\documentclass[12pt]{article}

\usepackage{amssymb, amsmath}
\usepackage{graphicx}
\usepackage{cite}
\usepackage{hyperref}
\def\be{\begin{equation}}
\def\ee{\end{equation}}
\def\bea{\begin{eqnarray}}
\def\eea{\end{eqnarray}}

\def\td{ \textrm{d}}

\def\nn{\nonumber}

\title{ {\bf Transverse deformations of extreme horizons}}

\author{Carmen Li\footnote{K.K.Li@sms.ed.ac.uk} \  and James Lucietti\footnote{j.lucietti@ed.ac.uk } \\ \\  \small \sl  School of Mathematics and Maxwell Institute for Mathematical Sciences, \\ \small \sl  University of Edinburgh, King's Buildings, Edinburgh, EH9 3FD, UK}

\date{}

\begin{document}

\maketitle

\begin{abstract}
We consider the inverse problem of determining all extreme black hole solutions to the Einstein equations with a prescribed near-horizon geometry. We investigate this problem by considering infinitesimal deformations of the near-horizon geometry along transverse null geodesics. We show that, up to a gauge transformation, the linearised Einstein equations reduce to an elliptic PDE for the extrinsic curvature of a cross-section of the horizon. We deduce that for a given near-horizon geometry there exists a finite dimensional moduli space of infinitesimal transverse deformations.   We then establish a uniqueness theorem for transverse deformations of the extreme Kerr horizon. In particular, we prove that the only smooth axisymmetric transverse deformation of the near-horizon geometry of the extreme Kerr black hole, such that cross-sections of the horizon are marginally trapped surfaces, corresponds to that of the extreme Kerr black hole. Furthermore, we determine all smooth and biaxisymmetric transverse deformations of the near-horizon geometry of the five-dimensional extreme Myers-Perry black hole with equal angular momenta. We find a three parameter family of solutions such that cross-sections of the horizon are marginally trapped, which is more general than the known black hole solutions. We discuss the possibility that they correspond to new five dimensional vacuum black holes.
\end{abstract}

\newpage

\tableofcontents

\newpage

\section{Introduction}

The classification of stationary black hole solutions to the Einstein equations is a major open problem in higher dimensional General Relativity~\cite{Emparan:2008eg}.  Besides being of intrinsic interest, it has numerous applications in modern studies of high energy physics. The key questions to be answered are: (1) What are the possible topologies and symmetries of black hole spacetimes? (2) What is the moduli space of black hole solutions with a given topology and symmetry?

The horizon topology theorem~\cite{Galloway:2005mf} and rigidity theorem~\cite{Hollands:2006rj} go some way to addressing question (1), revealing that cross-sections of the horizon must admit positive scalar curvature and that rotating black holes must be axisymmetric.  However, these results only provide necessary conditions that must be satisfied by black hole spacetimes. Furthermore, apart from topological censorship~\cite{Friedman:1993ty}, which guarantees the domain of outer communication (DOC) is simply connected~\cite{Chrusciel:1994tr, Galloway}, few general results towards constraining the topology of the DOC are known in higher dimensions.\footnote{Recent progress has been made in this direction~\cite{Andersson:2015sfa}.} Thus, it is far from clear what topologies and symmetries are actually realised by black hole solutions to Einstein's equations.

Question (2) is a much more refined question and would answer the fundamental question `What data characterises a black hole?'. Of course, for four dimensional asymptotically flat vacuum black holes, this is answered by the classic black hole uniqueness theorems revealing a very simple moduli space parameterised by the mass and angular momentum. In higher dimensions the situation is significantly more complicated and a rich moduli space is expected. General results in this direction are lacking, see~\cite{Hollands:2012xy} for a recent review of the state of the art.

Nevertheless, significant steps towards question (2) have been made for spacetimes with $\mathbb{R} \times U(1)^{D-3}$ symmetry. For $D=5$ this symmetry is compatible with asymptotic flatness and has allowed substantial progress in this context~\cite{Hollands:2007aj}. A black hole uniqueness theorem has been established which shows that in addition to the mass and angular momentum, certain data known as the rod structure must also be specified. The rod structure encodes the $U(1)^2$-action acting on the horizon and the DOC, together with certain associated geometric invariants. Thus there are an infinite number of possible rod structures. However, it is far from clear for what rod structures there exist  regular solutions to the Einstein equations and hence this result falls short of answering (2) in this case.  In fact the rod structure determines the topology of the horizon and the DOC and hence for this class the existence question would also answer question (1).  

Now, it might be tempting to expect that the only rod structures allowed are the ones of the known solutions.  For asymptotically flat black holes with connected horizons these are those of the Myers-Perry black hole and black rings.  However, recent work in supergravity shows that more complicated rod structures are possible. These correspond to black holes with non-trivial 2-cycles (bubbles) in the DOC~\cite{Kunduri:2014iga} and black holes with lens space topology horizons~\cite{Kunduri:2014kja} (black lenses). These examples are supersymmetric, although they raise the possibility of vacuum counterparts.

extreme black holes are of special interest due to their applications in quantum gravity. As is well known, these possess a precise notion of a near-horizon geometry that itself satisfies Einstein's equations. The classification of such near-horizon geometries is a much simpler -- yet still complicated -- task and substantial progress has been made~\cite{Kunduri:2013gce}.\footnote{The Einstein equations for a near-horizon geometry also arise from the restriction of the Einstein equations  for a spacetime to an extreme Killing/isolated horizon~\cite{Kunduri:2013gce, Lewandowski:2002ua}.} In particular this allows one to investigate possible horizon topologies and geometries, without reference to the parent black hole. Of course though, a major problem is to elevate such near-horizon classifications to black hole classifications. In particular, this requires answering the question:  (3) What is the set of extreme black hole solutions with a given near-horizon geometry? Equivalently, what spacetimes can an extreme horizon be embedded into such that it is the boundary of a black hole?

In fact, the classic black hole uniqueness theorem for the Kerr black hole has only recently been generalised to the extreme case~\cite{Amsel:2009et, Figueras:2009ci, Chrusciel:2010gq, Meinel:2011wu}. The new ingredient required is an understanding of the near-horizon geometry to derive the appropriate boundary conditions near the horizon. This has also been generalised to five-dimensional extreme vacuum black holes with $\mathbb{R}\times U(1)^2$ symmetry~\cite{Figueras:2009ci}. This reveals that it is only the $SO(2,1)\times U(1)^{D-3}$ near-horizon symmetry~\cite{Kunduri:2007vf} that enters the proof, rather than the explicit near-horizon geometry. The result is much like in the non-extreme case: the mass and angular momenta together with the rod structure suffice to uniquely determine the black hole. Therefore, the near-horizon geometry is also implicitly determined by this data. However, just like in the non-extreme case, the existence problem is not understood and therefore these works do not address question (3).

Question (3) has been investigated for certain supersymmetric black holes. This was first done for five dimensional minimal supergravity. It was shown that, under certain restrictive assumptions, the only asymptotically flat supersymmetric black hole with a near-horizon geometry locally isometric to that of the BMPV black hole is the BMPV black hole~\cite{Reall:2002bh}. This result makes use of global constraints on the spacetime from supersymmetry and asymptotic flatness. 

In this paper we will address question (3) directly. We should emphasise there is no guarantee that a given near-horizon geometry arises as the near-horizon limit of an extreme black hole. For example, there exists a trivial flat near-horizon geometry with toroidal topology, however, the horizon topology theorem rules out black holes with toroidal topology.  Furthermore, even if a black hole with a given near-horizon geometry exists, it may not be unique. For example, the near-horizon limits of the Myers-Perry black hole and slow rotating KK-black hole are isometric~\cite{Kunduri:2008rs}. Another example is given by (tensionless) vacuum black strings and black rings~\cite{Kunduri:2007vf}. These examples show that neither existence nor uniqueness of black hole solutions with a given near-horizon geometry is guaranteed.

Our strategy is to investigate question (3) {\it infinitesimally} near an extreme Killing horizon. The near-horizon limit of such a horizon is defined by taking a scaling limit of an affine parameter for transverse null geodesics. We will work to first order in this affine parameter. Hence our results are local to the horizon and do not employ any global constraints on the spacetime.  In fact, since the near-horizon limit of a solution to the Einstein equations is itself a solution, the problem reduces to studying the linearised Einstein equations for a particular class of perturbations of the near-horizon geometry. We term these perturbations {\it transverse deformations} of the near-horizon geometry (or extreme horizon). Gravitational perturbations of near-horizon geometries of extreme black holes have been studied before~\cite{Amsel:2009ev, Dias:2009ex, Durkee:2010ea, Hollands:2014lra}. These works considered general dynamical perturbations. The perturbations we study in this paper are non-dynamical and by construction are such that the perturbed spacetime still has an extreme Killing horizon.

We will show that once the appropriate diffeomorphism freedom is taken into account, the linearised Einstein equations for transverse deformations of an extreme horizon reduce to an elliptic PDE on a cross-sections $S$ of the horizon for the extrinsic curvature of $S$ (along the transverse null direction).  Therefore, assuming $S$ is compact allows us to deduce the following general result: {\it the moduli space of transverse deformations of an extreme horizon is finite dimensional}.   

We will also study the conditions for our transverse deformations to render $S$ a marginally trapped surfaces (MTS) and propose certain `extremality' conditions that must be met  for a MTS to correspond to the horizon of an extreme black hole. Interestingly, we will show that these extremality conditions are in fact guaranteed by the near-horizon symmetry theorems.

We will then analyse explicit solutions to our linearised Einstein equations. In particular, we are able to classify all axisymmetric transverse deformations of the extreme Kerr horizon. This reveals that the most general deformation which renders $S$ a MTS corresponds to that arising from the full extreme Kerr black hole itself. Thus we find a local version of the no-hair theorem emerging, logically distinct to the standard no-hair theorem which employs global assumptions such as asymptotic flatness. 

Perhaps more surprisingly, we also find it is possible to completely classify all biaxisymmetric transverse deformations of the five-dimensional extreme Myers-Perry horizon with equal angular momenta.  We find a three dimensional family of smooth solutions which render $S$ a MTS. A zero-dimensional subset correspond to the deformations of the full extreme Myers-Perry black hole with equal angular momenta (which possess enhanced $SU(2)\times U(1)$ rotational symmetry).  We are unaware of any known black hole solutions which exhibit  our general family of deformations. We discuss their possible interpretation in the Discussion.

The organisation of the paper is as follows. In section 2 we define the concept of an infinitesimal transverse deformation of an extreme horizon and derive the associated linearised Einstein equations. In section 3 we examine the consequences of imposing that our horizon is a MTS and propose our extremality conditions. In section 4 we determine all axisymmetric deformations of the known four dimensional near-horizon geometries. In section 5 we determine all $U(1)^2$-invariant deformations of the near-horizon geometry of a Myers-Perry black hole. We discuss our results in section 6. A number of the calculations are relegated to an Appendix.

\section{Transverse deformations of extreme horizons}

\subsection{Coordinates and gauge freedom}
\label{sec:coords}

Let $(M,g)$  be a $D$-dimensional spacetime containing a smooth \textit{degenerate} Killing horizon $\mathcal{H}$ of a complete Killing field $n$. In the neighbourhood of such a horizon we introduce Gaussian null coordinates and the associated near-horizon geometry, which we first recall. See e.g. the review~\cite{Kunduri:2013gce} for more details.

Let $S$ be a $D-2$ dimensional spacelike surface in $\mathcal{H}$ everywhere transverse to $n$, i.e. a cross section of $\mathcal{H}$.  Thus   $S_v = \psi_v [S]$, where $\psi_v$ is the 1-parameter group of isometries generated by $n$, defines a foliation of $\mathcal{H}$.  Now let  $(x^a)$ be a coordinate chart\footnote{Spacetime indices are denoted by Greek letters $\mu, \nu, \dots$, whereas cross-section $S$ indices are denoted by Latin letters $a,b,\dots$.} on $S$ with $a=1,...,D-2$ containing a point $p \in S$. We assign coordinates $(v,x^a)$ to the point in $S_v \subset \mathcal{H}$ a parameter value $v$ along the integral curve of $n$ passing through $p$, synchronised so that $v=0$ corresponds to $p$. This gives coordinates $(v,x^a)$ on a tubular neighbourhood of $p$ in $\mathcal{H}$ such that $n = \partial / \partial v$.  Now consider a point $q\in \mathcal{H}$ contained in a chart $(v,x^a)$. Let $\ell$ be the unique past directed null vector at $q$  orthogonal to $S_v$ such that $n \cdot \ell=1$. Assign coordinates $(v,r,x^a)$ to the point an affine parameter value $r$ along the integral curve of the null geodesic starting at $q$ with tangent $\ell$, synchronised so $r=0$ corresponds to $q$. This gives coordinates $(v,r,x^a)$ for a tubular neighbourhood of $q$ in $M$.  We define,
\be
n = \frac{\partial}{ \partial v} \; , \qquad \ell = \frac{\partial }{ \partial r} \; ,  \label{nl}
\ee 
everywhere in this chart, not just on $\mathcal{H}$. It is easily verified that the geodesic property of $\ell$ implies that $n \cdot \ell =1$ and $\ell \cdot \frac{\partial}{\partial x^a}=0$ for all $r$.  It can then be shown that the spacetime metric in these coordinates take the form
\be
g_{\mu \nu} \td x^\mu \td x^\nu = 2  \td v \left( \tfrac{1}{2} r^2 F(r, x) \td v + \td r + r h_a (r, x)  \td x^a \right) + \gamma_{ab} (r, x) \td x^a \td x^b \; .  \label{gnc}
\ee
Observe that degeneracy of the horizon is equivalent to $g_{vv}= O(r^2)$.  

The above construction defines a double foliation $S(v,r)$ of $M$ near $\mathcal{H}$, where $S(v,r)$ are surfaces of constant $(v,r)$, such that $S(0,0)=S$ agrees with our original cross-section. Thus $(x^a)$ also give coordinates on $S(v,r)$. The quantities $F, h_a, \gamma_{ab}$ transform as a function, 1-form and Riemannian metric under coordinate changes on $S(v,r)$ and so are globally defined on $S(v,r)$. It is important to realise that this coordinate system is unique up to a choice of cross-section $S$, the coordinates on $S$ and a constant rescaling $(v, r,x^a) \to ( v \lambda^{-1}, r \lambda, x^a)$.

For any degenerate Killing horizon as above there exists a well defined notion of a near-horizon geometry.
For any $\varepsilon >0$, consider the diffeomorphism $\phi_\varepsilon$ mapping a point with coordinates $(v,r,x^a)$ to one with coordinates $(v / \varepsilon,  \varepsilon r , x^a)$ and define the 1-parameter family of metrics $g_\varepsilon \equiv \phi^*_{\varepsilon} g$. Explicitly,
\be
g_\varepsilon = 2  \td v \left( \tfrac{1}{2} r^2  F( \varepsilon r, x) \td v + \td r + r h_a (\varepsilon r, x) \td x^a \right) + \gamma_{ab} (\varepsilon r, x) \td x^a \td x^b \;  .     \label{spacetime}
\ee
The near-horizon limit $\bar{g}$ -- also called the near-horizon geometry -- is defined as the $\varepsilon \rightarrow 0$ limit of $g_\varepsilon$. Smoothness of the metric functions guarantees it exists  and it is given by 
\be
\bar{g}= 2 \td v \left( \tfrac{1}{2 } r^2  F(x) \td v + \td r + r h_a (x) \td x^a \right) + \gamma_{ab} (x) \td x^a \td x^b  \; , \label{nhg}
\ee
where $F(x) = F(0, x)$, $h_a(x) = h_a(0, x)$ and $\gamma_{ab}(x) = \gamma_{ab} (0, x)$ are a function, 1-form and Riemannian metric on $S$.  Note that the near-horizon geometry is fully specified by the horizon data $(F, h_a, \gamma_{ab})$ on $S$. In particular,  $\gamma_{ab}$, is the induced metric on $S$.

Clearly if $g$ is an exact solution to the Einstein equation, its near-horizon geometry $\bar{g}$ is also an exact solution. Furthermore, it turns out that the Einstein equations $R_{\mu\nu}=\Lambda g_{\mu\nu}$ for a near-horizon geometry are equivalent to the following geometrical equations on $S$,
\bea
F &=& \tfrac{1}{2} h_a h^a - \tfrac{1}{2} D_a h^a +\Lambda  \; ,\label{F} \\
R_{ab} &=& \tfrac{1}{2} h_a h_b - D_{(a} h_{b)}  +\Lambda \gamma_{ab}\; ,\label{horeq}
\eea
where $D_a$ is the Levi-Civita connection and $R_{ab}$ the Ricci tensor of the metric $\gamma_{ab}$ on $S$. 

We may now state the precise problem we wish to study. That is, given a near-horizon geometry $\bar{g}$ that satisfies the Einstein equations,  determine all possible spacetimes (\ref{gnc}) that satisfy the Einstein equations with a near-horizon limit $\bar{g}$.
In other words, working in Gaussian null coordinates, given the data $(F,h_a, \gamma_{ab})$ on $S$, determine all possible $(F,h_a, \gamma_{ab})$ on $S(v,r)$.  Since all quantities are independent of $v$ this reduces to an ``evolution" problem in $r$: given the data $(F,h_a, \gamma_{ab})$ at $r=0$, determine all possible data $(F,h_a, \gamma_{ab})$ for $r>0$. In general this is a formidable problem. 

As a first step, in this paper, we will consider  this problem to first order in the affine parameter $r$.\footnote{Expansions of this type have been previously investigated for the more general isolated and dynamical trapping horizons~\cite{Booth:2012xm}.}  More precisely,  since smoothness allows us to Taylor expand the 1-parameter family of metrics $g_\varepsilon$ in $\varepsilon$, we will consider the problem to first order in  the parameter $\varepsilon$. Thus we define the first order {\it transverse deformation} of the near-horizon geometry or extreme horizon by,
\be
g^{(1)} = \left. \frac{\td g_\varepsilon}{\td\varepsilon}  \right|_{\varepsilon =0}  \; .
\ee
Explicitly, this is given by,
\be
g^{(1)}=r \left[  2  \td v \left( \tfrac{1}{2} r^2 F^{(1)}(x) \td v  + r h^{(1)}_a (x) \td x^a \right) +  \gamma^{(1)}_{ab} (x) \td x^a \td x^b \right]  \label{g1}
\ee
where,
\be
F^{(1)}(x) = \partial_r F(r,x) |_{r=0}  \qquad h^{(1)}_a  = \partial_r h_a(r,x) |_{r=0}, \qquad \gamma^{(1)}_{ab} = \partial_r 
\gamma_{ab}(r,x) |_{r=0}  \; .  \label{1storder}
\ee
Our problem therefore reduces to determining $g^{(1)}$ given $\bar{g}$, or equivalently, determining the first order data $(F^{(1)}, h^{(1)}_a, \gamma_{ab}^{(1)})$ given the horizon data $(F,h_a, \gamma_{ab})$ on $S$. 

The first order data has a direct geometrical interpretation. The null vectors $n$ and $\ell$ on $S$ used in the construction of Gaussian null coordinates above are both orthogonal to $S$.  The extrinsic curvatures of $S$ with respect to these null normals are defined at any point on $S$ by,
 \be
 \chi^{(\ell)}(X,Y) = X^\mu Y^\nu  \nabla_\mu \ell_\nu , \qquad \chi^{(n)}(X,Y) =X^\mu Y^\nu \nabla_\mu n_\nu  \; ,
 \ee
 where $\nabla$ is the spacetime Levi-Civita connection and $X,Y$ are tangent vectors to $S$. Explicitly evaluating these in Gaussian null coordinates (\ref{gnc}), we find
 \be
 \chi^{(\ell)}_{ab} = \tfrac{1}{2}  \gamma^{(1)}_{ab} , \qquad \chi^{(n)}_{ab}=0  \; .   \label{extrinsic}
 \ee
The first relation thus provides an interpretation for the first order quantity $\gamma^{(1)}_{ab}$ (\ref{1storder}). The second relation is of course a consequence of the fact that $S$ is a cross-section of a Killing horizon of the Killing field $n$.  

It is important to realise that although Gaussian null coordinates are unique up to the choice of $S$ and coordinates on $S$, there is an ambiguity in the first order expansion. This corresponds to diffeomorphisms which preserve the form of the metric to first order in $\varepsilon$. Clearly this is a weaker demand that preserving the form of the metric to all orders in $\varepsilon$.  To see this, let $\chi_\varepsilon$ be a 1-parameter family of diffeomorphisms generated by a vector field $\xi$.  Then, $(M,g_{\varepsilon})$ and $(M,\tilde{g}_\varepsilon=\chi^*_\varepsilon g_{\varepsilon})$ are diffeomorphic spacetimes.  Thus $g^{(1)}$ and
\be
\tilde{g}^{(1)} =  \left. \frac{\td (\chi^*_\varepsilon g_{\varepsilon})}{\td\varepsilon}  \right|_{\varepsilon =0}  = g^{(1)} - \mathcal{L}_\xi \bar{g}
\ee
represent equivalent transverse deformations.  Demanding that $\tilde{g}^{(1)}$, and hence $\mathcal{L}_\xi \bar{g}$, takes the same form as (\ref{g1}) restricts the possible vector fields $\xi$. Writing $(\mathcal{L}_\xi \bar{g})_{\mu\nu}= 2 \bar{\nabla}_{(\mu} \xi_{\nu )}$, where $\bar{\nabla}$ is the Levi-Civita connection of the near-horizon geometry $\bar{g}$, we see that determining $\xi$ reduces to a calculation in the near-horizon geometry.

It is convenient to work in a null orthonormal frame for the near-horizon geometry. We define this by $e^\mu$ where $\mu=+, -, a$ and $a=1, \dots, D-2$, so that $\bar{g} = 2 e^+ e^- + e^a e^a$, where
\bea
e^+ = \td v \qquad e^- = \td r + r h_a \td x^a + \tfrac{1}{2} r^2 F \td v, \qquad e^a = \hat{e}^a  \; , \label{basis}
\eea
where $\hat{e}^a$ is an orthonormal frame for $(S, \gamma_{ab})$.\footnote{To avoid a proliferation of indices we will use Greek indices $\mu, \nu, \dots$ to denote both coordinate and orthonormal frame indices on $M$, and Latin letters $a,b, \dots$ to denote both coordinate and orthonormal frame indices on $S$.}  The spin connection and curvature in this basis are given in~\cite{Kunduri:2013gce}. In this basis, the form of the first order deformation (\ref{g1}) is preserved if and only if
\be
\bar{\nabla}_{(-} \xi_{\mu)}=0 \;, \qquad \bar{\nabla}_{(a} \xi_{b)} = O(r), \qquad \bar{\nabla}_{(a} \xi_{+)}= O(r^2), \qquad \bar{\nabla}_{(+} \xi_{+)}= O(r^3).
\ee
It is easily checked that the most general vector field with this property is  given by
\be
\xi= \tfrac{1}{2}f e^- + \left[ r h^aw_a+ \tfrac{1}{4} r^2 ( F f - h^{ a} D_a f) \right] e^+ + \left[w_a- \tfrac{1}{2} r (D_a f ) \right]e^a \;,
\ee
where $f$ is a function on $S$ and $w_a$ is a Killing field of $\gamma_{ab}$ which preserves the horizon data $\mathcal{L}_w F=\mathcal{L}_w h_a=0$. The diffeomorphisms generated by $w_a$ are simply isometries of the near-horizon geometry tangent to $S$ and do not affect the first order data, so we discard these. The remaining diffeomorphisms are thus generated by a single function $f$ on $S$. A straightforward calculation then reveals,
\begin{eqnarray}
\tilde{\gamma}^{(1)}_{ab} &=&\gamma^{(1)}_{ab} + D_a D_b f - h_{(a} D_{b)} f  \label{gautransgamma}\\
\nn \tilde{h}^{(1)}_a & = & h^{(1)}_a  - \tfrac{1}{2} F D_a f - \tfrac{1}{4} (D_a h_{b})  D^b f - \tfrac{1}{4} h_a h_b D^b f + \tfrac{1}{2} (D_b h_a )D^b f + \tfrac{1}{4} h_b D_a D^b f \label{gautransh} \\
\tilde{F}^{(1)} & = & F^{(1)} + \tfrac{1}{2} (D^a f) \left( D_a F - h_a F \right) \label{gautransf}   \nn  
\end{eqnarray}
are the gauge transformation rules for the first order data.  Observe that since the metric components must remain smooth we require $f$ to be a smooth function on $S$.  

We emphasise that this gauge freedom is merely an artefact of our first order expansion. As already mentioned above, Gaussian null coordinates for the full metric are always unique once a cross-section $S$ and coordinates on $S$ are chosen.  Thus, geometrical properties of $S$ such as its extrinsic curvature (\ref{extrinsic}) lead to unambiguous expressions in this chart. The gauge freedom reflects the artificial redundancy that appears by only requiring the spacetime metric to be in Gaussian null coordinates to first order in the affine parameter $r$.

Before moving on, we mention a useful fact which will be useful later when examining explicit deformations.  Let $K^a$ generate a symmetry of the horizon data and suppose the deformation is also invariant under this symmetry, i.e., $\mathcal{L}_K \gamma^{(1)}_{ab}=0$.  The gauge transformations that preserve this condition must obey $D^2 (\mathcal{L}_K f) - h^a D_a (\mathcal{L}_K f)=0$ and hence if $S$ is compact $\mathcal{L}_K f$ is a constant. Integrating over $S$ we get $\int_S \mathcal{L}_K f = \int_S K^a D_a f = -\int_S (D_a K^a) f= 0$, so the constant $\mathcal{L}_K f$ must in fact vanish. Thus the gauge transformation function $f$ must also be invariant under $K$.

\subsection{Linearised Einstein equations}

We will consider spacetimes $(M,g)$ of the form (\ref{gnc}) that are solutions to the Einstein equations $R_{\mu\nu}=\Lambda g_{\mu\nu}$. Thus, the diffeomorphic spacetimes $(M,g_{\varepsilon})$ given by (\ref{spacetime}) are also a solution to the Einstein equation. In particular, as  observed above, it follows that the near-horizon geometry $\bar{g}$ is also a  solution. Define
\be
R^{(1)}_{\mu\nu} \equiv \left. \frac{\td }{\td\varepsilon} R_{\mu\nu}(g_\varepsilon)  \right|_{\varepsilon =0}= \bar{\Delta}_L g^{(1)}_{\mu \nu} +\bar{\nabla}_{(\mu } v_{\nu)}   \label{Ric1}
 ,
\ee
where $v_\mu = \bar{\nabla}_\nu g^{(1)~\nu}_{~ ~ \mu} - \frac{1}{2} \partial_\mu ( g^{(1)}_{\rho \sigma} \bar{g}^{\rho \sigma})$ and 
\begin{equation}
\bar{\Delta}_L g^{(1)}_{\mu \nu} = - \tfrac{1}{2} \bar{\nabla}^{2} g^{(1)}_{\mu \nu} - \bar{R}_{\mu ~ \nu}^{~ \kappa ~ \lambda} g^{(1)}_{\kappa \lambda} + \bar{R}_{(\mu }^{~ ~ \kappa} g^{(1)}_{\nu) \kappa} 
\end{equation}
is the Lichnerowicz operator of the background $\bar{g}$ and $\bar{R}_{\mu\nu\rho\sigma}$ is the Riemann tensor of $\bar{g}$. Then, then linearised Einstein equation is simply
\be
R^{(1)}_{\mu\nu} = \Lambda g^{(1)}_{\mu\nu} \; .  \label{linE}
\ee
Therefore, the transverse deformation $g^{(1)}$ obeys the linearised Einstein equation in the background of a near-horizon geometry $\bar{g}$.

The explicit components of the linearised Ricci tensor $R^{(1)}_{\mu\nu}$ are listed in our null-orthonormal basis (\ref{basis}) in the Appendix. These all reduce to covariant equations on $S$. The $-a$ component of (\ref{linE}) may be solved directly for $h^{(1)}_a$ giving,
\be
 h^{(1)}_a = \tfrac{1}{2} h^{b} \gamma^{(1)}_{ab} - \tfrac{1}{2} {D}^b \gamma^{(1)}_{ab} + \tfrac{1}{2} {D}_a  \gamma^{(1)}- \tfrac{1}{4} h^{}_a  \gamma^{(1)}   \; , \label{h1}
 \ee
 where $\gamma^{(1)} \equiv  \gamma^{ab} \gamma^{(1)}_{ab}$.
Similarly, the $+-$ component of (\ref{linE}) can be solved for $F^{(1)}$ resulting in,
 \begin{eqnarray}
\nonumber F^{(1)} &=&  h^{a} h_a^{(1)} - \tfrac{1}{3} {D}^a h_a^{(1)} - \tfrac{1}{3} h^{a} h^{b} \gamma^{(1)}_{ab}  + \tfrac{1}{6} h^{(a} {D}^{b)} \gamma^{(1)}_{ab}  \\
& &+ \tfrac{1}{6}\left( {D}^{(a} h^{b)} \right) \gamma^{(1)}_{ab}  - \tfrac{1}{6} F  \gamma^{(1)}  - \tfrac{1}{12} h^{a} \left( {D}_a  \gamma^{(1)}   - h^{}_a  \gamma^{(1)}  \right)   \label{f1eqn} \; . 
\end{eqnarray} 
These equations determine the first order data  $h^{(1)}_a, F^{(1)}$ in terms of $\gamma^{(1)}_{ab}$ (and the horizon data).  The problem thus reduces to determining $\gamma^{(1)}_{ab}$. The $ab$ component of (\ref{linE}) can be simplified using (\ref{h1}) and the horizon equation (\ref{F}), resulting in a PDE on $S$ for  $\gamma^{(1)}_{ab}$,
\bea
  0 &=&   \Delta_L \gamma^{(1)}_{ab} + \tfrac{1}{2} {D}_{(a} 		{D}_{b)}\gamma^{(1)}  + \tfrac{3}{2} h^{c} {D}_c \gamma^{(1)}_{ab} - \tfrac{3}{2} h^{}_{(a} {D}_{b)}\gamma^{(1)}  - h^{c} {D}_{(a}  \gamma^{(1)}_{b)c} + h^{}_{(a} {D}^c  \gamma^{(1)}_{b)c} \nn \\
  &-&   \tfrac{1}{2} h^{2} \gamma^{(1)}_{ab} + \tfrac{1}{2} h^{}_a h^{}_b \gamma^{(1)}  + \left( {D}_{(a} {h^{}}^c\right) \gamma^{(1)}_{b)c}    - \left( {D}^c h^{}_{(a}\right) \gamma^{(1)}_{b)c}  \;  ,  \label{g1abeqn}
\eea
where 
\be
 \Delta_L \gamma^{(1)}_{ab} =- \tfrac{1}{2} D^2  \gamma^{(1)}_{ab} + {R}_{(a}^{~ ~ c} \gamma^{(1)}_{b)c} - {R}_{a~b~}^{~c~d}\gamma^{(1)}_{cd}
 \ee 
 is the Lichnerowicz operator and $R_{abcd}$ is the Riemann tensor of  the cross-section $(S, \gamma_{ab})$.  Observe that (\ref{g1abeqn}) is automatically traceless; this is related to the gauge freedom in our problem as we discuss below. Also note that the cosmological constant has cancelled out. It can be checked that the remaining components of the linearised Einstein equations impose no further constraint (see Appendix).

To summarise, we have shown that the linearised Einstein equation reduces to solving the PDE (\ref{g1abeqn}) defined on the cross-section $S$ for the extrinsic curvature (\ref{extrinsic}), with the remaining first order data $h^{(1)}_a, F^{(1)}$ then determined {\it algebraically} by (\ref{h1}) and (\ref{f1eqn}). As observed above, the first order data is only defined up to the gauge transformations (\ref{gautransgamma}). It is a straightforward, yet tedious, exercise to check that (\ref{h1}, \ref{f1eqn}, \ref{g1abeqn}) are indeed invariant under these gauge transformations.

Let us now count the degrees of freedom in our linearised problem. We have a symmetric tensor $\gamma^{(1)}_{ab}$ on $S$ subject to the gauge transformation (\ref{gautransgamma}). This gives $\tfrac{1}{2} (D-2)(D-1) - 1$ independent components. On the other hand, the linearised Einstein equation (\ref{g1abeqn}), which is symmetric and automatically traceless, gives $\tfrac{1}{2} (D-2)(D-1) - 1$ independent equations. Therefore the problem is neither over or under determined and one expects at most a discrete set of solutions.  More precisely, using the gauge freedom one may fix the trace $\gamma^{(1)}$, so then (\ref{g1abeqn}) is an elliptic equation for the traceless part of $\gamma^{(1)}_{ab}$. From standard Fredholm theory we deduce the following result. \\

\noindent {\bf Proposition.}
Consider a near-horizon geometry with a compact cross-section $S$ and horizon data $(F,h_a,\gamma_{ab})$. The moduli space of solutions $\gamma^{(1)}_{ab}$ to the linearised Einstein equation \eqref{g1abeqn}, modulo the gauge transformation \eqref{gautransgamma}, is finite dimensional.  \\

This is one of the main results of this paper. Observe that it allows for the possibility of no solution, a unique solution, or multiple solutions parameterised by a finite number of parameters.

It is worth noting that for $D=3$, so $S$ is 1-dimensional, the linearised Einstein equation (\ref{g1abeqn}) is trivially satisfied. This agrees with the analysis of~\cite{Li:2013pra} where all 3d Einstein metrics of the form (\ref{gnc}) where determined exactly, as we confirm in the Appendix. In fact, we deduce that in 3d the first order analysis is essentially enough to determine the full solution.

 \section{Marginally trapped surfaces and extremality}
 \label{sec:MTS}
 
 We are interested in deformations of near-horizon geometries corresponding to black hole spacetimes. There are several notions of quasi-local horizons which are intended to capture the idea of a black hole, see~\cite{Booth:2005qc} for a comprehensive review.  We will require that the cross-section $S$ is a marginally trapped surface (MTS), or equivalently, that the Killing horizon $\mathcal{H}$ is a marginally trapped tube.   
 
 The null vectors $n$ and $\ell$ on $S$ used in the construction of Gaussian null coordinates (\ref{nl}) are both orthogonal to $S$.  The expansions of these null vectors are obtained by tracing the corresponding extrinsic curvatures (\ref{extrinsic}) with respect to the induced metric on $S$. Thus, we get
 \be
 \theta_\ell  = \tfrac{1}{2} \gamma^{(1)} \; , \qquad \theta_n = 0  \; .
 \ee
 Therefore, $S$ is a marginally trapped surface iff $\theta_\ell>0$ everywhere on $S$.  Thus we require 
 \be
 \gamma^{(1)}>0
 \ee
 everywhere on $S$.
 
Recall there is a gauge freedom in the linear deformations introduced above given by the transformations (\ref{gautransgamma}). In particular these imply that
 \be
 \gamma^{(1)} \to \gamma^{(1)} + D^2 f - h^a D_a f
 \ee
 for any smooth function $f$ on $S$. Clearly these transformations need not preserve the sign of $\theta_\ell$. Therefore, we deduce that to linear order in our deformation parameter $\varepsilon$, we are unable to impose the condition that $S$ is a marginally trapped surface in a manner that is invariant under our gauge transformations. This suggests that a first order analysis is insufficient to determine whether a deformation renders $S$ a MTS. In fact, this is not quite so.
 
 Remarkably, there exists a weaker condition which must be satisfied if $S$ is a MTS, which is gauge invariant.  Firstly, we exploit the existence of a particular decomposition of the 1-form $h_a$ on $S$, proven in~\cite{Lucietti:2012sf}. This states that there exists a unique (up to scale) positive function $\Gamma$ on $S$ such that,
 \be
 h = \Gamma^{-1}h' - \td\log \Gamma \; ,  \label{hdecomp}
 \ee
 and $D_a h'^a=0$.\footnote{Note this is not the same as the Hodge decomposition of $h_a$, which is sometimes used in this context.}  It follows that
 \be
\Gamma \gamma^{(1)}  \to \Gamma \gamma^{(1)} + D^a (\Gamma D_a f - h'_a f)   \; ,
\ee
and therefore the constant
\be
\Theta \equiv \tfrac{1}{2} \int_S \Gamma \gamma^{(1)}   \label{H}
\ee
is gauge invariant. Therefore, a necessary condition for $S$ to be a MTS is $\Theta>0$.  This allows us to deduce, for instance, that if $S$ is a MTS there is no gauge in which $\gamma^{(1)}=0$ everywhere on $S$.
 
 As is well known, the notion of a marginally trapped surface is not sufficiently restrictive to capture the idea of a black hole. For this we require input on how $\theta_n$ changes as we deform $S$ to a surface just inside or outside the horizon. Typically this is implemented by requiring that there exist trapped surfaces just inside the horizon.   Concretely, this is equivalent to the existence of a scaling of the null fields $(\ell, n) \to ( e^\lambda \ell, e^{-\lambda} n)$, for some function $\lambda$ on $S$, such that $\mathcal{L}_\ell \theta_k >0$ on $S$, where $k$ is the unique null-extension of the null vector $n$ on $S$ into $M$ which is orthogonal to the deformation surfaces and normalised by $\ell \cdot k =1$~\cite{Hayward:1993wb}. In the context of marginally outer trapped surfaces (MOTS) this is often referred to as the `stability' condition~\cite{Andersson:2007fh}. It guarantees that just inside the horizon $\theta_k<0$ everywhere on the corresponding deformation surfaces, thus ensuring they are trapped (note by continuity we must also have $\theta_\ell>0$ on these surfaces, since we have assumed $S$ is a MTS).
 
 However, for extreme black holes this is not the correct criterion since they do not contain any trapped surfaces, see e.g.~\cite{Booth:2007wu}. Instead, we require that there exists a scaling of the null fields such that
 \be
 \mathcal{L}_\ell \theta_k=0, \qquad \mathcal{L}_\ell \mathcal{L}_\ell \theta_k >0   \label{extreme}
 \ee 
 everywhere on $S$. This guarantees that just inside and outside the horizon $\theta_k>0$, i.e., there exist untrapped surfaces both inside and outside the horizon. The first criterion in (\ref{extreme}) was derived in~\cite{Booth:2007wu} and more recently~\cite{Mars:2012sb}. The second criterion in (\ref{extreme}) we propose as a `stability' condition for degenerate MTS.\footnote{This condition was mentioned as a possible `degenerate' case in~\cite{Hayward:1993wb}.} We take both of these criteria as necessary for a degenerate horizon to be the event horizon of an extreme black hole.
 
In Gaussian null coordinates, a natural choice for the deformation surfaces are the surfaces $S(v,r)$ introduced above. By construction  the null vector $\ell$ is always orthogonal to $S(v,r)$. On the other hand, the unique null vector $k$ orthogonal to $S(v,r)$ that satisfies $\ell \cdot k=1$ on $S(v,r)$ and coincides with $n$ at $r=0$, is
 \be
 k = \partial_v - r h^a \partial_a + \tfrac{1}{2} r^2 (h^2-F) \partial_r   \; .
 \ee
 The induced metric on $S(v,r)$ is $\gamma_{ab}$.  The extrinsic curvatures of $S(v,r)$ with respect to these null normals at any point on $S(v,r)$ are given by
 \be
  \chi^{(\ell)}(X,Y) = X^\mu Y^\nu \nabla_\mu \ell_\nu  , \qquad \chi^{(k)} (X,Y) = X^\mu Y^\nu \nabla_\mu k_\nu   \; ,
 \ee
 where $X,Y$ are now tangent vectors to $S(v,r)$.
 A short calculation shows that in Gaussian null coordinates (\ref{gnc}) these are given by,
 \bea
 \chi^{(k)}_{ab} &=& \tfrac{1}{4} r^2 (h^2-F) \partial_r \gamma_{ab} - r D_{(a} h_{b)}  \; ,\\
 \chi^{(\ell)}_{ab} &=& \tfrac{1}{2} \partial_r \gamma_{ab}  \; ,
 \eea
 where $D_a$ is the Levi-Civita connection of $\gamma_{ab}$.
 Observe these reduce to the expression (\ref{extrinsic}) on $S$, as they should. The expansion are obtained by taking the trace of these with respect to $\gamma_{ab}$, 
 \bea
 \theta_k &=& \tfrac{1}{4} r^2 (h^2-F) \gamma^{ab} \partial_r \gamma_{ab} - r D_a h^a  \; ,\\
 \theta_\ell &=& \tfrac{1}{2} \gamma^{ab} \partial_r \gamma_{ab}  \; .
 \eea
 We may now evaluate the first variation of $\theta_k$ in the transverse direction $\ell$. This is given by\footnote{This agrees with the expression (3.10) in~\cite{Racz:2008tf} derived for solutions to the Einstein equations, upon using (\ref{horeq}) and noting their $\beta_a =-h_a$. In deriving our expression we made no use of Einstein's equation.}
 \be
 \mathcal{L}_\ell \theta_k |_{S} = - D_a h^a |_{r=0}  \; .   \label{1stvar}
 \ee
 Typically, this is non-vanishing with an indefinite sign. For example, this is so for the near-horizon geometry of extreme Kerr, see section (\ref{sec:kerr}).  This shows that the deformation surfaces $S(v,r)$ defined by Gaussian null coordinates are not in general the ones appropriate for imposing the criterion (\ref{extreme}). As discussed above, to obtain the correct deformations one must boost the null fields by an appropriate function on $S$. Fortunately, there is a straightforward way to do this.
 
 Consider a boost $\ell' = \Gamma \ell$, where $\Gamma$ is a positive function on $S$. It is easily verified that the geodesics of $\ell'$ are also affinely parameterised. In particular $\ell' = \partial / \partial r'$ where $r' = \Gamma^{-1} r$ is the new affine parameter.  Now change from Gaussian null coordinates to new coordinates defined by $(v', r',x'^a)= (v, \Gamma^{-1} r, x^a)$. The spacetime metric becomes,
 \be
 g= 2 \td v' \left( \tfrac{1}{2} r'^2 F' \td v' + \Gamma \td r' + r' h'_a \td x'^a \right) + \gamma'_{ab} \td x'^a \td x'^b  
 \ee
 where $F'= \Gamma^2 F$, $h'_a = \Gamma h_a + \partial_a \Gamma$ (note this coincides with the decomposition of $h_a$ on $S$ used above) and $\gamma'_{ab}=\gamma_{ab}$.  Define the surfaces $S(v',r')$ as the surfaces of constant $(v',r')$. Observe that $S(0,0)=S$ as before, although in general $S(v',r')$ are not the same surfaces as $S(v,r)$. The two null vectors which are orthogonal to $S(v',r')$ and coincide with $\ell'$ and $n$ on $S$, are $\ell'$ and
 \be
 k' =  \partial_{v'} - r h'^a \partial'_a + \tfrac{1}{2\Gamma} r'^2 (h'^2-F') \partial_{r'}   \; ,
\ee
which are normalised as $\ell' \cdot k' = \Gamma$. By a similar calculation as above, the expansions of $k', \ell'$ on $S(v',r')$ are then,
\bea
 \theta_{k'} &=& \tfrac{1}{4} r'^2  \Gamma^{-1} (h'^2-F') \gamma'^{ab} \partial_{r'} \gamma'_{ab} - r' D'_a h'^a \\
 \theta_{\ell'} &=& \tfrac{1}{2} \gamma'^{ab} \partial_{r'} \gamma'_{ab}  \; .
 \eea
Observe that $\theta_{\ell'} = \Gamma \theta_\ell$. Thus, as one would expect, the condition that $S$ is a MTS remains $\theta_{k'}=0$ and $\theta_{\ell'}>0$. 

We now compute the transverse variations of $\theta_{k'}$ along $S(v',r')$. Again, the first variation of $\theta_{k'}$ on $S$ is given by (\ref{1stvar}) with all quantities replaced by their primed versions. Therefore, if we choose $\Gamma$ as in the decomposition (\ref{hdecomp}) we deduce
\be
\mathcal{L}_{\ell'} \theta_{k'} |_{S} = - D'_a h'^a |_{r'=0} =0  \; ,  \label{1stvarzero}
\ee
as required. We may now compute the second variation of $\theta_{k'}$ along our deformation surfaces $S(v',r')$. We find,
\be
\mathcal{L}_{\ell'} \mathcal{L}_{\ell'} \theta_{k'} |_S = - \tfrac{1}{2} A \gamma'^{(1)} - D'^a ( h'^{(1)}_a + \tfrac{1}{2} h'_a \gamma'^{(1)})  \; , \label{2ndvar}
\ee
where $\gamma_{ab}'^{(1)} = \partial_{r'} \gamma'_{ab}|_{r'=0}$ etc, we have defined the function \be
A=\Gamma^{-1} (F'-h'^2)|_{r'=0},   \label{A}
\ee
and we have used the special choice of $\Gamma$ which ensures (\ref{1stvarzero}).   We now investigate the sign of the second variation.

Observe that the quantity $A$ is a function on $S$ which only depends on the near-horizon data. Its significance is revealed by writing the near-horizon geometry in terms of it,
\be
\bar{g} = \Gamma ( r'^2 A \td v'^2 + 2 \td v' \td r' )+ \gamma_{ab}'(\td x'^a+ r h'^a \td v)(\td x'^b+ r h'^b \td v)   \; .
\ee
The near-horizon symmetry enhancement theorems establish that in a wide class of theories which includes vacuum gravity, a non-trivial near-horizon geometry must be such that $A=A_0<0$ is a negative constant and $h'^a$ is a Killing vector field~\cite{Kunduri:2007vf, Figueras:2008qh, Lucietti:2012sa, Li:2013gca}. This result has been established under various assumptions of rotational symmetry and implies the isometry group of the near-horizon geometry is at least $SO(2,1)\times U(1)$, see the review~\cite{Kunduri:2013gce} for more details.   Assuming this symmetry enhancement result and noting that for a MTS we have $\gamma'^{(1)}= \Gamma \gamma^{(1)}>0$, we see that the first term in (\ref{2ndvar}) is positive. Therefore, we have shown the following result. \\

\noindent {\bf Proposition.} Let $\mathcal{H}$ be a degenerate Killing horizon as above, with a compact cross-section $S$ that is a MTS. Suppose that in the near-horizon geometry $A=A_0<0$ is a constant and $h'^a$ is a Killing field on $S$, so that it possesses $SO(2,1)\times U(1)$ symmetry. Then,
\be
\int_S \mathcal{L}_{\ell'} \mathcal{L}_{\ell'} \theta_{k'}   = -A_0 \Theta  >0  \; ,
\ee
where $\Theta= \int_S \theta_{\ell'}$ is the same constant as defined by (\ref{H}). \\

This shows that a weaker version of our criteria (\ref{extreme}) is guaranteed for a large class of extreme horizons under the assumption they are MTS.  In fact, this is the best one can do for linear deformations. As already observed, for our linear deformations we may not impose the MTS condition in a gauge invariant manner. However, we found that a weaker condition $\Theta>0$ could be imposed. Similarly, the positivity of the second variation (\ref{2ndvar}) is not preserved by the gauge transformations (\ref{gautransgamma}) and therefore cannot be imposed to linear order. However, the weaker condition that $\int_S \mathcal{L}_{\ell'} \mathcal{L}_{\ell'} \theta_{k'} >0$ is gauge invariant. It is interesting that we have found that its positivity is a consequence of near-horizon symmetry enhancement.

\section{Four dimensional solutions}

We will now investigate the moduli space of transverse deformations to the known four dimensional near-horizon geometries.  We will first summarise what is known about the classification of such objects~\cite{Kunduri:2013gce}. For simplicity we will focus on vacuum solutions and so we set $\Lambda=0$ henceforth.  Assuming compactness of the cross-section $S$, it can be shown that $S$ must be either topologically toroidal $T^2$ or spherical $S^2$.  The toroidal case can be completely determined and is given by a flat near-horizon geometry with a flat metric on $T^2$.  The spherical case can be completely determined under the additional assumption of axisymmetry and corresponds to the near-horizon geometry of the extreme Kerr black hole. 

\subsection{Deformations of toroidal horizon}
\label{sec:toroidal}

As a warm up, let us first consider the possible deformations of the flat $T^2$ extreme horizon.   
For this case, the horizon data $(F,h_a, \gamma_{ab})$ on $S \cong T^2$ is trivial,
\be
R_{abcd}=0 \; , \qquad h_a=0 \; , \qquad F=0  \; .
\ee
Without loss of generality we will scale our torus so that it has unit volume.
The equation for transverse deformations (\ref{g1abeqn}) and the allowed gauge transformations (\ref{gautransgamma}) are simply,
\be
- D^2 \gamma_{ab}^{(1)} + D_{a} D_{b} \gamma^{(1)}=0  \; ,  \label{flatdeform}
\ee
and 
\be
\gamma^{(1)}_{ab} \to \gamma^{(1)}_{ab} + D_a D_b f  \; ,
\ee
respectively.  

We may define a 1-form
\be
X_a = \epsilon^{bc} D_{b} \gamma^{(1)}_{ca}
\ee
where $\epsilon_{ab}$ is the volume form on $T^2$. It is easily seen this is invariant under the gauge transformations. Taking a derivative of (\ref{flatdeform}) it then follows that
\be
D^2 X_a=0  \; .
\ee
Compactness then implies that the 1-forms are in fact covariantly constant $D_a X_b=0$.  Furthermore,
\be
\int_{S} X^a X_a = \int_S X^a D_b (\epsilon^{bc} \gamma^{(1)}_{ca}) = \int_S  D_b (X^a \epsilon^{bc} \gamma^{(1)}_{ca}) = 0
\ee
where in  the second equality we have used the fact that $X^a$ is covariantly constant. Hence the gauge invariant 1-form must vanish,
\be
X_a=0  \; .
\ee
However, this does not imply the general solution is gauge equivalent to the trivial solution $\gamma^{(1)}_{ab}=0$.

A convenient choice of gauge is found by demanding that the trace
\be
\tilde{\gamma}^{(1)} = \gamma^{(1)} +D^2 f
\ee
is a constant. The existence of such a gauge thus reduces to the existence of solutions to the Poisson equation on compact $S$. It is well known that a unique (up to an additive constant) smooth solution exists if and only if $\int_S (\gamma^{(1)} - \tilde{\gamma}^{(1)}) =0$.  Thus, we require,
\be
\tilde{\gamma}^{(1)} = \int_S \gamma^{(1)} = 2\Theta  \; ,
\ee
where in the second equality we have written this constant in terms of the gauge invariant quantity (\ref{H}) (note we can take $\Gamma=1$ in this case).  In this gauge (\ref{flatdeform}) becomes $D^2 \tilde{\gamma}^{(1)}_{ab}=0$ and therefore by compactness we deduce $D_c \tilde{\gamma}^{(1)}_{ab}=0$.  

To summarise, we have shown that there exists a gauge in which $\gamma^{(1)}_{ab}$ is covariantly constant. Observe that from (\ref{h1}) and (\ref{f1eqn}) it follows that in this gauge the rest of the first order data is trivial $h^{(1)}_a=0$ and $F^{(1)}=0$. Of course, there is nothing here which forbids solutions with $\Theta>0$ so that $S$ is a MTS. However, since the function (\ref{A}) vanishes for the near-horizon geometry, from (\ref{2ndvar}) we have  $\mathcal{L}_\ell \mathcal{L}_\ell \theta_k |_S =0$ so that our second extremality criterion (\ref{extreme}) is not satisfied. By the horizon topology theorems we already know that this near-horizon geometry cannot arise as a near-horizon limit of an extreme black hole.  This is consistent with our proposal that (\ref{extreme}) is the correct extremality criterion for a MTS to be an extreme black hole.

It is worth mentioning the vacuum plane wave solutions
\be
ds^2 = 2 \td v \td r + \gamma_{ab}(r) \td x^a \td x^b  \label{planewave}
\ee
where $\gamma_{ab}(r)$ is a matrix satisfying  $2\gamma^{ab} \ddot{\gamma}_{ab} + \dot{\gamma}^{ab}\dot{\gamma}_{ab}=0$ and $\cdot$ indicate $r$-derivatives. Clearly, by periodically identifying the coordinates $(x^a)$, these can be interpreted as spacetimes with a near-horizon geometry with toroidal horizon. Observe these solutions have $F=0, h_a=0$ for all $r \neq 0$. This is of course consistent with our first order analysis above. Observe that these include solutions with $\gamma^{(1)}>0$ so that $S$ is a MTS. However, as mentioned above they violate our second extremality condition (\ref{extreme}).

\subsection{Deformations of extreme Kerr horizon}
\label{sec:kerr}

The extreme Kerr horizon data is given by~\cite{Lewandowski:2002ua, Kunduri:2008rs},
\bea
\gamma_{ab} dx^a dx^b &=& a^2 \frac{1+x^2}{1-x^2} \td x^2+ 4 a^2 \frac{(1-x^2)}{1+x^2} \td\phi^2  \label{kerrhor} \\ 
h_a dx^a &=& \frac{4(1-x^2)}{(1+x^2)^2} \td\phi  - \frac{2x}{1+x^2} \td x \, \\ F &=&  \frac{3-6x^2-x^4}{a^2 (1+x^2)^3}  \; ,
\eea
where $a>0$ is a constant and the coordinate ranges are $-1<x<1$ and $\phi \sim \phi +2\pi$. The horizon metric and data extends smoothly to the endpoints $x = \pm 1$ giving a metric on $S^2$. To see this explicitly, write $x= \pm (1- \epsilon^2)$ and expand for small $\epsilon$. Then, the horizon metric approaches $\sim 2a^2 ( \td\epsilon^2+ \epsilon^2 \td\phi^2)$ so that identifying $\phi$ with period $2\pi$ avoids any conical singularity. The endpoints $x=\pm 1$ are fixed points of the axisymmetry Killing field $m=\partial_\phi$.

We now consider  transverse deformations to this extreme horizon, that is, {\it smooth} solutions $\gamma^{(1)}_{ab}$ to (\ref{g1abeqn}) in the background of the extreme Kerr horizon. To render the problem tractable, we will assume axisymmetry so that $\mathcal{L}_{m} \gamma^{(1)}_{ab}=0$. By the argument at the end of section (\ref{sec:coords}) the gauge transformation functions that preserve this condition must be axisymmetric $\mathcal{L}_m f=0$.\footnote{Another way to see that the constant $\mathcal{L}_m f$ must vanish is that otherwise $f$ would be monotonic along orbits of $m$, which would violate periodicity. }

 In the $(x,\phi)$ chart  the components of $\gamma^{(1)}_{ab}$ only  depend only on the coordinate $x$. Hence the deformation equation (\ref{g1abeqn}) reduces to ODEs. Explicitly, the three components of $\gamma^{(1)}_{ab}$ obey two independent ODEs,
\begin{eqnarray}
\nonumber 0&=&-4 x^2 (x^2 -3)^2(1+x^2) \gamma^{(1)}_{\phi \phi} + (1 - x^2) \left\lbrace 16 x^3 (x^2 -1) \gamma^{(1)}_{x \phi} + 6 x (x^2-2)(1+x^2)^2 \gamma^{(1)\prime}_{\phi \phi} \right. \\
& & \left. + (x^2 -1) \left[ 8 (x^4 -1) \gamma^{(1)\prime}_{x \phi} - 8 x (x^2 -1) \gamma^{(1)\prime}_{xx} + (1 + x^2)^3 \gamma^{(1)\prime\prime}_{\phi \phi}  \right] \right\rbrace   \label{EKg1eqnxx} \\
\nonumber 0 &=& 8 x (x^2-3)(1+x^2) \gamma^{(1)}_{\phi \phi} + 8 (- x^6 + 3 x^4 + x^2 + 1) \gamma^{(1)}_{x \phi} + (x^2 -1) \left\lbrace 5 (1+ x^2)^2 \gamma^{(1)\prime}_{\phi \phi}  \right. \\
& & \left. + 4 x (1+x^2)(1-3x^2) \gamma^{(1)\prime}_{x \phi} - 2 (x^2 - 1) \left[ 2 (1 - x^2) \gamma^{(1)\prime}_{xx} + (1+ x^2)^2 \gamma^{(1)\prime\prime}_{x \phi} \right] \right\rbrace \; ,  \label{EKg1eqnxp}
\end{eqnarray} 
where the prime denotes derivative with respect to $x$. These correspond to the $\phi\phi$ and $x\phi$ components respectively (the $xx$ can be written in terms of these as a consequence of the tracelessness of (\ref{g1abeqn})). The gauge transformations \eqref{gautransgamma} are,
\begin{eqnarray}
\gamma^{(1)}_{xx} & \rightarrow & \gamma^{(1)}_{xx} - \frac{2x^3}{1-x^4} f' + f''   \label{EKgautransxx}   \\
\gamma^{(1)}_{x \phi} & \rightarrow & \gamma^{(1)}_{x \phi} - \frac{2 (1- x^2)}{(1+x^2)^2} f'    \label{EKgautransxp}   \\
\gamma^{(1)}_{\phi \phi} & \rightarrow & \gamma^{(1)}_{\phi \phi} - \frac{8 x (1- x^2)}{(1+x^2)^3} f'    \label{EKgautranspp}    \; , 
\end{eqnarray}
where $f=f(x)$ as argued above.

Smoothness of $\gamma_{ab}^{(1)}$ translates to boundary conditions on its components at the endpoints $x =\pm 1$. 
Transforming to $x= \pm (1-\epsilon^2)$ as above and converting the polar coordinates $(\epsilon, \phi)$ near each pole to cartesian coordinates, it can be easily shown that the components of $\gamma^{(1)}_{ab}$ must satisfy
\be
 \gamma^{(1)}_{x\phi} = O(1-x^2), \qquad \gamma^{(1)}_{\phi\phi}= O(1-x^2), \qquad \gamma^{(1)}_{xx} = \frac{\gamma^{(1)}_{\phi\phi}}{(1-x^2)^2} + O(1)
\ee
as $x \to \pm 1$ where all $O$ terms are smooth functions of $x$ up to and including the endpoints. Thus the $O$ terms extend to smooth axisymmetric functions on $S^2$.

We will now rewrite this system in terms of gauge invariant variables. It is straightforward to see that the quantity 
\be
X = 4 x \gamma^{(1)}_{x \phi} - (1+ x^2) \gamma^{(1)}_{\phi \phi}  \label{EKX}
\ee
is gauge invariant. Furthermore, from our boundary conditions it is clear that $X$ is a smooth function on $S^2$ which vanishes at the poles $x =\pm 1$. Less obviously the quantity 
\bea
\nn Y & =&  x^2 \left(1+x^2\right)^3 \left(1-x^2\right)^2  \gamma^{(1)\prime\prime}_{\phi \phi}-2 x(1+x^2)^2(1+2x-2x^2)(1-2x-2x^2) (1-x^2)\gamma^{(1)\prime}_{\phi \phi} \\
& & +2 (1+x^2)(1-4x^2+26x^4-20x^6+5x^8) \gamma^{(1)}_{\phi \phi}-8 x^3 \left(x^2-1\right)^3 \gamma^{(1)\prime}_{xx}\label{EKY} \; , 
\eea
is also a gauge invariant smooth function on $S^2$ which vanishes at the poles \footnote{An easy way to see smoothness is to write $Y$  in terms of the globally defined vector field $(1-x^2)\partial_x$ on $S^2$ which vanishes at $x=\pm 1$.}. Rewriting \eqref{EKg1eqnxx} and \eqref{EKg1eqnxp} in terms of these gauge invariant variables simplifies the ODEs to,
\bea
0 &=& 2 x \left(x^2+1\right) \left(x^2-1\right)^3 X'+2 \left(x^4+1\right) \left(x^2-1\right)^2 X+Y \label{EKg11}\\
\nn 0 &=& x^2 \left(x^4-1\right)^2 X''-2 x (1-x^4)(1-x^2+2x^4) X'\\
& & -2 \left(5 x^6+x^4+3 x^2-1\right) X+Y \label{EKg12} \; . 
\eea
Subtracting them gives  a remarkably simple second order ODE for just $X$,
\be
0 = 2 (x^2+1) X +(1-x^2) \left[2x X' -  (1-x^2) X'' \right]   \; .
\ee
The general solution to this is simply 
\be
X = \frac{A x (x^2 -3)+ B}{1-x^2}  \; ,
\ee
where $A,B$ are constants of integration.  We could substitute back to  determine $Y$. We have thus fully solved for the local form of the solution, in terms of our gauge invariant variables.

We may now impose our boundary conditions. Recall that we require $X$ to be a smooth function of $x$ which vanishes at $x=\pm 1$. This immediately forces the constants $A=B=0$. Therefore we have shown that the only solutions compatible with our boundary conditions is the trivial one
\be
X= Y=0  \; .
\ee
To determine the explicit components $\gamma_{ab}^{(1)}$ we must now invert  \eqref{EKX} and \eqref{EKY} for $X=Y=0$. We find,
\be
\gamma^{(1)}_{xx}  = c + \frac{(1+x^2)\left[ 2 x (2 x^2 -3) \gamma^{(1)}_{x \phi} - (1-x^4) \gamma^{(1)\prime}_{x \phi} \right]}{2 (1-x^2)^2}  \; ,
\qquad \qquad 
\gamma^{(1)}_{\phi \phi} = \frac{4 x \gamma^{(1)}_{x \phi}}{1+ x^2}  \; ,\label{kerrg1absol1}
\ee
where $c$ is a constant. Therefore the general solution is specified by an arbitrary function $\gamma^{(1)}_{x \phi} (x)$ and a constant $c$.

To reveal the interpretation of the constant $c$, we compute the trace of $\gamma^{(1)}_{ab}$, which we may write as,
\be
\gamma^{(1)} = \frac{ c(1-x^2) - \tfrac{1}{2} [ (1+x^2)^2 \gamma^{(1)}_{x\phi}]'}{a^2(1+x^2)}   \;.
\ee
The function $\Gamma$, defined by the decomposition (\ref{hdecomp}), can be taken to be~\cite{Kunduri:2008rs},
\be
\Gamma= \tfrac{1}{2}(1+x^2)  \; .
\ee
Therefore, we find that the invariant (\ref{H}) is simply
\be
\Theta= \frac{8 \pi c}{3}  \; .
\ee
Notice the arbitrary function $\gamma^{(1)}_{x\phi}$ does not appear in this expression as a consequence of the boundary conditions $\gamma^{(1)}_{x\phi}=0$ at $x =\pm 1$. This is consistent with the fact that $\Theta$, unlike $\gamma^{(1)}_{x\phi}$, is gauge invariant. Therefore, $c$ sets the scale of the deformation. For $S$ to be a MTS we thus require
\be
c>0 \; .
\ee
Of course, it is really only the sign of $c$ that has physical meaning. 

We now compare our general solution to that obtained by writing the full extreme Kerr black hole in Gaussian null coordinates. This calculation cannot be done exactly, instead one must work order by order in the coordinate $r$. This is not a problem since all we need are the first order corrections to the near-horizon geometry. The calculation is relegated to the Appendix. We find
\be
\gamma^{(1) EK}_{xx} = \frac{4a}{1-x^4}, \qquad \gamma^{(1) EK}_{x\phi} = \frac{4a x(1-x^2)}{(1+x^2)^2}, \qquad \gamma^{(1) EK}_{\phi\phi}  =\frac{16 a x^2(1-x^2)}{(1+x^2)^3}  \; .  \label{kerrgam1}
\ee
It is easily verified that the gauge invariant functions $X,Y$ vanish for this deformation.

We will now show that our general solution is in fact gauge equivalent to the extreme Kerr data. It is convenient to define a smooth function $g(x)$ by
\be
\gamma^{(1)}_{x\phi} = \frac{g(x) (1-x^2)}{(1+x^2)^2}  \; .
\ee
In terms of this the gauge transformation for $\gamma^{(1)}_{x\phi}$ simply reads $g(x)\to  g(x)- 2f'$.  Thus, let us choose $f$ such that in the new gauge $\gamma^{(1)}_{x\phi}$ agrees with a multiple of that of extreme Kerr. Denoting the data in the new gauge with tildes this means $\tilde{\gamma}^{(1)}_{x\phi} = \Omega \gamma^{(1) EK}_{x\phi}$  where $\Omega>0$ is a constant. Since $g^{EK}(x)= 4a x$, the condition for this is,
\be
f'= \tfrac{1}{2} g(x) - 2ax \Omega  \; ,  \label{fkerr}
\ee
which defines $f$ up to a an irrelevant additive constant.  Using this $f$ it is clear that we also have $\tilde{\gamma}^{(1)}_{\phi\phi} = \Omega \gamma^{(1) EK}_{\phi\phi}$. The final component is more non-trivial. Using (\ref{fkerr}) one finds that all dependence on $g(x)$ in the new gauge cancels, leaving,
\be
\tilde{\gamma}_{xx}^{(1)} = c- 2 a\Omega+ \frac{4a \Omega x^4}{1-x^4}  \; .
\ee
Thus, if we choose $f$ such that $\Omega= c/(6a)$, we deduce that $\tilde{\gamma}^{(1)}_{xx} = \Omega \gamma^{(1) EK}_{xx}$. Therefore, we have shown that our general solution with $c>0$ is gauge equivalent to a positive multiple of the extreme Kerr data, $\tilde{\gamma}^{(1)}_{ab} = \Omega \gamma^{(1) EK}_{ab}$.

To summarise, we have established the following  `local uniqueness' theorem for transverse deformations of the extreme Kerr horizon. \\

\noindent {\bf Theorem.} (Uniqueness of transverse deformations of extreme Kerr horizon). The most general smooth axisymmetric solution to (\ref{g1abeqn}) for the extreme Kerr horizon such that $S$ is a MTS, is gauge equivalent to (a positive multiple of) the first order data of the extreme Kerr black hole. \\

Thus we find that there is a unique (up to scale) solution to the linearised Einstein equations. Hence the moduli space of infinitesimal axisymmetric deformations in this case is zero-dimensional, in line with our general result.
We emphasise this is logically distinct to the standard no hair theorem for extreme Kerr. No input about the global structure of the spacetime, such as asymptotic flatness, is used.  

\section{Five dimensional solutions}

In this section we will investigate deformations of known five dimensional near-horizon geometries. Again, for simplicity we will only consider vacuum solution so we set $\Lambda=0$. First we recall that the classification of vacuum near-horizon geometries, with compact $S$, invariant under $U(1)^2$ symmetry has been fully solved~\cite{Kunduri:2008rs}. It turns out the non-trivial solutions are locally isometric to the near-horizon geometries of the extreme black ring/string, the extreme Myers-Perry black holes (or slow rotating extreme KK black holes), or the fast rotating extreme KK black hole.

For simplicity, we will only consider the simplest non-trivial near-horizon geometry in this context. This is the near-horizon limit of the extreme Myers-Perry black hole with equal angular momenta. This solution enjoys an enhancement of rotational symmetry to $SU(2)\times U(1)$.  Cross-sections $S$ are homogeneously squashed $S^3$. In fact, it turns out this the most general vacuum homogeneous near-horizon geometry~\cite{Kunduri:2013gce}. Its horizon data can be written as,
\bea
\gamma &=& \frac{4}{k^2} ( \td\psi+ x \td\phi)^2 + \frac{2}{k^2} \left( \frac{\td x^2}{1-x^2} +(1-x^2)\td\phi^2 \right)  \label{homohor} \\
h &=& 2 (\td\psi+ x \td\phi) \\
F &=& \tfrac{1}{2} k^2
\eea
where $k>0$ is a constant. The coordinate ranges are $-1<x<1$ and $\Delta \phi=2\pi $ and $\Delta \psi = 4\pi$ (setting $x=\cos\theta$ gives standard Euler coordinates). The metric extends smoothly to the endpoints $x=1$ and $x=-1$, which correspond to the fixed points of the rotational Killing fields $m_1 = \partial_\phi - \partial_\psi$ and $m_2=\partial_\phi + \partial_\psi$ respectively.  

We will need to perform a global analysis of various geometric quantities on $S$. For this it is convenient to use coordinates $(\phi_1, \phi_2)$ adapted to the Killing fields such that $m_i = \partial / \partial \phi_i$ for $i=1,2$.  These are given by 
\be
\phi= \phi_1+\phi_2, \qquad \psi= \phi_2-\phi_1.  \label{phii}
\ee
The horizon metric in these coordinates is 
\be
\gamma = \frac{2}{k^2} \left( \frac{\td x^2}{1-x^2}  + (1-x)(3-x) \td\phi_1^2+ (1+x)(3+x) \td\phi_2^2 - 2 (1-x^2) \td\phi_1 \td\phi_2 \right)  \; .
\ee
Setting $x= 1-\epsilon^2$ and expanding for small $\epsilon$ the metric approaches $ \sim 4 k^{-2}( \td\epsilon^2+ \epsilon^2 \td\phi_1^2+ 8 \td\phi_2^2  - 4 \epsilon^2 \td\phi_1 \td\phi_2)$. Converting the polar coords $(\epsilon, \phi_1)$ to cartesian coords it is easily seen the metric smoothly approaches $\mathbb{R}^2\times S^1$ provided $\phi_1 \sim \phi_1+2\pi$. A similar calculation confirms smoothness at $x=-1$ provided $\phi_2 \sim \phi_2+ 2\pi$.  It will be useful to note that 
\be
D=(1-x^2) \frac{\partial}{\partial x} \label{Svec}
\ee
is a smooth globally defined  vector field on $S^3$ and vanishes at $x=\pm 1$ (to see this convert to cartesian coords near $x=\pm 1$.)

Let us now consider possible deformations $\gamma_{ab}^{(1)}$ in this background. Due to the symmetry of the near-horizon geometry, there are a number of possible symmetry assumptions for this deformation. In particular, we may assume that $\gamma^{(1)}_{ab}$ is invariant under any subgroup of the isometry group of $(S,\gamma_{ab})$. For example, we may assume $\gamma^{(1)}_{ab}$ preserves the full $SU(2)\times U(1)$ symmetry, i.e. homogeneous deformations. This problem is algebraic and easy to solve. More interestingly,  we may instead assume the deformation only preserves $U(1)^2$ symmetry. This problem reduces to a system of ODEs which remarkably can be fully solved, as we show below.

\subsection{Homogeneous deformations}
\label{sec:homodef}

The most general deformation invariant under the $SU(2)\times U(1)$ symmetry of the near-horizon geometry is
\be
\gamma^{(1)}_{ab}\td x^a \td x^b = c \left( \frac{\td x^2}{1-x^2} +(1-x^2)\td\phi^2 \right) +\tilde{c}\, (\td\psi+ x \td\phi)^2  \label{homodef}
\ee
where $c,\tilde{c}$ are constants.  To preserve this symmetry, the argument at the end of section (\ref{sec:coords}) shows that the gauge transformation function $f$ must be invariant under the homogeneous symmetry and hence must be constant. Therefore, for homogeneous deformations $\gamma^{(1)}_{ab}$ is in fact a gauge invariant quantity.

Substituting into (\ref{g1abeqn}) reveals that 
\be
\tilde{c}=0
\ee
although $c$ may be any constant. Computing the trace gives 
\be
\gamma^{(1)} = c k^2   \; .
\ee
Thus $S$ is a MTS if and only if
\be
c>0  \; .
\ee
Observe that for homogeneous deformations the MTS condition may be fully implemented since $\gamma^{(1)}$ is a gauge invariant quantity as observed above.

We will now compare to the known solutions. The extreme MP black hole with equal angular momenta has a near-horizon geometry of the form (\ref{homohor}). As shown in the Appendix its parameter is given by $k= 2/a$ and the first order transverse deformation arising from the full black hole solution is of the above form with $c= a$. Also, as shown in the Appendix, the extreme KK black hole with zero angular momentum has an isometric near-horizon geometry with $k = \frac{2 \sqrt{p+q}}{p \sqrt{q}}$. In this case, the first order deformation is also of the above form  with $c = \sqrt{pq}$.  In fact, since the precise value of the scale of $c$ is not determined by our first order analysis, it is impossible to distinguish these two black hole solutions at this order.

\subsection{$U(1)^2$-invariant deformations}
\label{sec:toricdef}

We now consider deformations $\gamma^{(1)}_{ab}$ that are invariant only under the $U(1)^2$ of the near-horizon geometry. In the coordinate $(x,\phi, \psi)$ introduced above, this means the components $\gamma^{(1)}_{ab}$ are only functions of $x$. To preserve this symmetry, the argument at the end of section (\ref{sec:coords}) shows that the gauge transformation function $f$ must also be $U(1)^2$-invariant. Hence $f$ may only depend on $x$. 

The explicit gauge transformations are then,
\bea
\nn && \gamma^{(1)}_{xx} \rightarrow \gamma^{(1)}_{xx} - \frac{ x f'}{1- x^2} + f'' \; , \qquad  \gamma^{(1)}_{x \phi} \rightarrow  \gamma^{(1)}_{x\phi}-{x f'}\;, \qquad  \gamma^{(1)}_{x \psi} \rightarrow \gamma^{(1)}_{x\psi} -  f' \; , \\
&& \gamma^{(1)}_{\phi \phi} \rightarrow \gamma^{(1)}_{\phi\phi} +x (1-x^2) f' \; ,  \qquad  \gamma^{(1)}_{\phi \psi} \rightarrow \gamma^{(1)}_{\phi\psi}+  (1-x^2) f' \; , \qquad \gamma^{(1)}_{\psi \psi} \rightarrow \gamma^{(1)}_{\psi \psi}   \; .
 \label{5dmpgautrans}
\eea
Thus $\gamma^{(1)}_{\psi\psi}$ is a gauge invariant quantity.  It is easily checked that,
\bea
\nn W(x) &=&  (1-x^2)^3 \gamma^{(1)\prime\prime}_{x \psi}  - x (1-x^2)^2 \gamma^{(1)\prime}_{x \psi} -(1- x^4) \gamma^{(1)}_{x \psi} +  (1-x^2)^3 \gamma^{(1)\prime}_{xx}\\
\nn X(x) &=&  \gamma^{(1)}_{\phi \phi} + x (1-x^2) \gamma^{(1)}_{x \psi} \\
\nn Y(x) &=&  x (1-x^2) \gamma^{(1)}_{x \psi}  - (1-x^2)  \gamma^{(1)}_{x \phi} \\ 
Z(x) &=& \gamma^{(1)}_{\phi \psi} + (1- x^2) \gamma^{(1)}_{x \psi} \label{5dgauinv} \;  ,
\eea
are also all gauge invariant and smooth on $S^3$. To find $W$ we wrote the linearised Einstein equations (which we know are gauge invariant) in terms of the gauge invariant quantities $X,Y,Z$.  It is straightforward to see that these variables are all smooth on $S^3$ by writing them in terms of the globally defined vector field (\ref{Svec}). For example, $Y$ can be written as the smooth invariant $Y=x \gamma^{(1)}(D, \partial_\psi) - \gamma^{(1)}(D, \partial_\phi)$, which furthermore shows that it must vanish at $x=\pm 1$. Similarly, $W$ can be written as smooth invariant which vanishes at $x=\pm 1$. Clearly, the other gauge invariant variables $\gamma^{(1)}_{\psi\psi}, X, Z$ can also be written as smooth invariants, although these need not vanish at $x=\pm 1$.  Thus we have five smooth gauge invariant functions $\gamma^{(1)}_{\psi\psi}, W, X, Y,Z$ and we find that we can write all components of the linearised Einstein equations in terms of these. This makes sense since $\gamma^{(1)}_{ab}$ has 6 components and the gauge transformation removes one degree of freedom.   

We now give the linearised Einstein equation (\ref{g1abeqn}) written in terms of these gauge invariant variables. Due to the traceless of (\ref{g1abeqn}) only 5 of the 6 components are independent. We discard the $xx$ component since this is the most complicated. The $x\phi$, $x \psi$, $\phi \phi$, $\phi \psi$ and $\psi \psi$ components are given by
\bea
\nonumber 0 &=& - 3 x \left(1-x^4\right)   \gamma^{(1)\prime}_{\psi \psi} + 2 \left(x^4-6 x^2+1\right) \gamma^{(1)}_{\psi \psi}-2 x W - 6 x \left(1- x^2\right)  X'-8 x^2 X\\
& & +2 \left(1-x^2\right)^2 Y'' +4  \left(1-x^2\right) Y-2 \left(7 x^4-8 x^2+1\right) Z'+8 x \left(x^2+1\right)  Z \label{xphieqn}\\
\nonumber 0 &= & \left(x^4+4 x^2-5\right) \gamma^{(1)\prime}_{\psi \psi} +4 x \left(x^2-3\right) \gamma^{(1)}_{\psi \psi} -2 W - 6 \left(1-x^2\right) X'-8 x X \\
& & - 4 \left(1-x^2\right) Y'+12 x \left(1-x^2\right) Z'+4 \left(x^2+3\right) Z \label{xpsieqn} \\
\nonumber 0 &=&  - \left(1-x^4\right) x \gamma^{(1)\prime}_{\psi\psi} -2 \left(x^4+4 x^2-1\right) \gamma^{(1)}_{\psi \psi} -2 x W  +2 \left(1-x^2\right)^2 X''  - 6  x\left(1-x^2\right)  X' \\
& & + 4 \left(1-3 x^2\right) X - 8 x \left(x^2-1\right)  Y'-4 \left(2 x^4-3 x^2+1\right) Z'+16 x^3 Z   \label{phiphieqn} \\
\nonumber 0 &=& - \left(x^4-4 x^2+3\right) \gamma^{(1)\prime}_{\psi\psi} - 8 x \gamma^{(1)}_{\psi \psi} -2 W - 8 x X - 6 \left(1-x^2\right) X'+4 \left(1-x^2\right) Y' \\
& & +2 \left(1-x^2\right)^2 Z''+ 4 x \left(1-x^2\right) Z'+8 \left(1+x^2\right) Z \label{phipsieqn} \\
0 &=& - \left(1-x^2\right) \gamma^{(1)\prime\prime}_{\psi \psi}-2  x \gamma^{(1)\prime}_{\psi\psi} + 4 Z' \label{psipsieqn} \; , 
\eea
respectively.  Thus, we have a system of 5 coupled second order ODEs for the 5 gauge invariant variables. Remarkably, this system of ODEs can be completely integrated in terms of elementary functions, as we now show.

First we observe that no derivatives of $W$ appear in any of the equations and therefore it may be solved for algebraically and eliminated from our system.  In particular rearranging (\ref{xpsieqn}) gives
\bea
W &= & \tfrac{1}{2} \left(x^4+4 x^2-5\right) \gamma^{(1)\prime}_{\psi\psi} +2x \left(x^2-3\right) \gamma^{(1)}_{\psi \psi}  - 3 \left(1-x^2\right) X'-4 x X \nn \\
& & - 2 \left(1-x^2\right) Y'+6 x \left(1-x^2\right) Z'+2 \left(x^2+3\right) Z   \; .  \label{W}
\eea
Substituting this expression for $W$ into (\ref{xphieqn}), (\ref{phiphieqn}) and (\ref{phipsieqn}) leads to welcome simplifications resulting in,
\bea
\nonumber 0 &=& x \left(1-x^2\right) \gamma^{(1)\prime}_{\psi\psi}+ \left(x^2+1\right) \gamma^{(1)}_{\psi \psi} +\left(1- x^2\right) Y''+2  x Y'+2  Y \\
& & - \left(1-x^2\right) Z'- 2 x Z \label{xphieqn1} \\
\nonumber 0 &=& 2 x \gamma^{(1)\prime}_{\psi\psi} +  \left(3 x^2+1\right) \gamma^{(1)}_{\psi \psi}+\left(1- x^2\right) X''+2 X(x)+6 x Y'\\
& &  - 2 \left(1+x^2\right) Z'-6 x Z \label{phiphieqn1} \\
0 &=& \left(x^2+1\right) \gamma^{(1)\prime}_{\psi\psi}+2 x \gamma^{(1)}_{\psi \psi} +4 Y' +  \left(1-x^2\right) Z''-4  x Z' - 2 Z \label{phipsieqn1}   \; ,
\eea
respectively. Thus, we are left with a system of four ODEs (\ref{xphieqn1}), (\ref{phiphieqn1}), (\ref{phipsieqn1}) and (\ref{psipsieqn}), for four variables $\gamma^{(1)}_{\psi\psi}, X,Y,Z$. In fact, observe that the three ODEs (\ref{xphieqn1}), (\ref{phipsieqn1}) and (\ref{psipsieqn}) do not involve $X$ and give a closed system for the variables $\gamma^{(1)}_{\psi\psi}, Y,Z$. Given $\gamma^{(1)}_{\psi\psi}, Y,Z$, the remaining equation (\ref{phiphieqn1}) can then then be used to determine $X$.

Thus let us consider the ODE system (\ref{xphieqn1}), (\ref{phipsieqn1}) and (\ref{psipsieqn}). In fact, (\ref{phipsieqn1}) is a total derivative. Integrating this gives
\be
\gamma^{(1)}_{\psi \psi} =  \frac{\left(x^2-1\right) Z'(x)+2 x Z(x) -4 Y(x)+c'}{ x^2+1} \; , \label{5du1sqgammapsipsi}
\ee
where $c'$ is a constant. This allows us to eliminate $\gamma^{(1)}_{\psi \psi}$ from (\ref{xphieqn1}) and (\ref{psipsieqn})  resulting in an ODE system for just $Y,Z$, which we refrain from writing down (note (\ref{psipsieqn}) becomes third order). Remarkably, adding \eqref{psipsieqn} to $4 \times$\eqref{xphieqn1}, with $\gamma^{(1)}_{\psi\psi}$ eliminated using (\ref{5du1sqgammapsipsi}), results in a dramatic simplification giving an ODE for just $Z$,
\be
0 =  \left(1- x^2\right)^2 Z''' - 4 (1- x^2) (x Z'' - Z') - 2 c'   \; .
\ee
This ODE is easily integrated in terms of elementary functions and generically has logarithmic singularities at $x=\pm 1$.  The general solution which is smooth at both $x=\pm 1$ requires,
\be
c'=0 \; ,
\ee
and is simply,
\be
Z = ax^2+b  \; ,  \label{Zsol}
\ee
where $a,b$ are constants. Now, substituting back into (\ref{xphieqn1}) gives a second order ODE for $Y$. The unique solution $Y$ to this ODE which vanishes at the endpoints $x=\pm 1$ is,
\be
Y= \tfrac{1}{8}(b - 3a) x (1-x^2)^2 \label{Ysol}
\ee
Now, (\ref{5du1sqgammapsipsi}) determines
\be
\gamma^{(1)}_{\psi\psi} =\tfrac{1}{2} x [ a(3x^2-1) + b(3-x^2) ] \; .   \label{psipsisol}
\ee
To summarise, we have now found that the general smooth solution to the ODE system (\ref{xphieqn1}), (\ref{phipsieqn1}) and (\ref{psipsieqn}), such that $Y=0$ at $x=\pm 1$, is given by (\ref{Zsol}, \ref{Ysol}, \ref{psipsisol}) and is parameterised by two constants $a,b$.

We may now substitute (\ref{Zsol}, \ref{Ysol}, \ref{psipsisol}) into the remaining ODE (\ref{phiphieqn1}) to get a second order ODE for $X$. This is also integrated in terms of elementary functions and the general solution for $X$ has logarithmic singularities at $x=\pm 1$.  The general solution which is smooth at $x= \pm 1$ is given by
\be
X = 
c (1-x^2)+\tfrac{1}{8} x [ a (1+10x^2-3x^4)+ b( 9 - 2x^2+x^4)]\label{Xsol}  \; ,
\ee
where $c$ is an integration constant. Finally,  substituting into (\ref{W}) determines
\be
W= 
 \tfrac{1}{8}(1-x^2)[ a( 13-50x^2+21x^4)+ b( -11+10x^2-7x^4) +16 c x]\label{Wsol}
\ee
Observe that this automatically obeys the required boundary conditions that $W=0$ at $x=\pm 1$.
We have now fully solved our original ODE system.

To summarise, we have found the general solution to the ODE system (\ref{xphieqn}), (\ref{xpsieqn}), (\ref{phiphieqn}), (\ref{phipsieqn}), (\ref{psipsieqn}) that is smooth for all $x\in [-1,1]$ and obeys the boundary conditions $Y=W=0$ at $x=\pm 1$. It is given by (\ref{Zsol}), (\ref{Ysol}), (\ref{psipsisol}), (\ref{Xsol}) and (\ref{Wsol}) parameterised by three constants $a,b,c$.   To orientate ourselves, it is worth recording that the deformations with enhanced $SU(2)\times U(1)$ symmetry discussed in the previous section, are given by the $a=b=0$ solution, 
\be
\gamma^{(1)}_{\psi\psi} = 0, \qquad W = 2c x (1-x^2), \qquad X=c(1-x^2), \qquad Y=Z=0
\ee
where $c$ is the same constant appearing in (\ref{homodef}). The solutions with non-zero $a,b$ thus represent a more general family of deformations which preserve only $U(1)^2$ symmetry.

Let us now reconstruct the deformation $\gamma^{(1)}_{ab}$ from our gauge invariant variables. Inverting (\ref{5dgauinv}) for our general solution, we find,
\bea
\gamma^{(1)}_{xx}&=& d+ \tfrac{7}{8} (3a-b)x + \frac{2c- x(b+2a)}{2(1-x^2)} + \frac{ x \gamma^{(1)}_{x\psi}}{1-x^2}- \gamma^{(1)\prime}_{x\psi}\\
\gamma^{(1)}_{x\phi}&=& x \gamma^{(1)}_{x\psi}-\tfrac{1}{8}(b - 3a) x (1-x^2) \\
\gamma^{(1)}_{\phi\phi} & = & c (1-x^2)+\tfrac{1}{8} x [ a (1+10x^2-3x^4)+ b( 9 - 2x^2+x^4)] - x(1-x^2)\gamma^{(1)}_{x\psi} \\
\gamma^{(1)}_{\phi \psi} &=&a x^2+b   - (1-x^2) \gamma^{(1)}_{x\psi}  \; ,
\eea
where $d$ is a new integration constant. Thus, it is parameterised by the four constants $a,b,c,d$ and an arbitrary function $\gamma^{(1)}_{x\psi}(x)$.  It is worth noting that smoothness of $\gamma^{(1)}_{ab}$ on $S$ implies $\gamma^{(1)}_{x\psi}(x)$ is a smooth function of $x$ (although not necessarily vanishing). We now must verify that our solution $\gamma_{ab}^{(1)}$ does indeed define a smooth tensor on $S^3$. 

To do this, it is simplest to convert to the $(\phi_1, \phi_2)$ coordinates (\ref{phii}).  We find,
\bea
\gamma^{(1)}_{x1} &=& -(1-x)\left[  \gamma^{(1)}_{x\psi}+\tfrac{1}{8}(b - 3a) x (1+x) \right] \label{soln} \\
\gamma^{(1)}_{x2} &=& (1+x) \left[ \gamma^{(1)}_{x\psi}-\tfrac{1}{8}(b - 3a) x (1-x) \right] \nn  \\
\gamma^{(1)}_{12} &=& (1-x^2) \left[ c+ \tfrac{1}{8} a ( 5+3x^2)- \tfrac{1}{8} b x (3+x^2) - x \gamma^{(1)}_{x\psi} \right] \nn \\
\gamma^{(1)}_{11} &=& \tfrac{1}{8} (1-x) \left[   8 c(1+x) + a x( - 3-19x+3x^2+3x^3)  \right. \nn \\ && \left. \qquad \qquad+ b( - 16+ 5x+5 x^2- x^3-x^4) +8(2-x)(1+x)\gamma^{(1)}_{x\psi}\right] \nn \\
\gamma^{(1)}_{22} &=& \tfrac{1}{8} (1+x) \left[ 8 c (1-x)  + a x( -3 + 19x + 3x^2-3x^3) \right. \nn \\  &&\left. \qquad \qquad +b( 16+5x-5x^2-x^3+x^4)  -8(2+x)(1-x)\gamma^{(1)}_{x\psi}\right]  \; .\nn
\eea
Thus, our deformation takes the form,
\bea
\nn \gamma^{(1)}_{ab} \td x^a \td x^b &=&  \mathcal{A}(x) \frac{\td x^2}{1-x^2} + \mathcal{B}(x) (1-x) \td \phi_1^2 + \mathcal{C}(x) (1+x) \td \phi_2^2 + \mathcal{D}(x) (1-x^2) \td \phi_1 \td \phi_2 \\
 & &  + \mathcal{E}(x) (1-x) \td x \td  \phi_1 + \mathcal{F}(x) (1+x) \td x \td \phi_2  \; , \label{fosmooth}
\eea
where the functions $\mathcal{A}(x),\mathcal{B}(x)...$ are all smooth functions on $x\in [-1,1]$. By converting to cartesian coords near each endpoint $x=\pm 1$, it is easy to see that all the off-diagonal terms are smooth at $x=\pm 1$. On the other hand, the diagonal terms lead to a conical singularity at $x=1$ and $x=-1$ unless $\mathcal{B}(1)= 2 \mathcal{A}(1)$ and $\mathcal{C}(-1)= 2 \mathcal{A}(-1)$, respectively. Remarkably, reading off the explicit form for $\mathcal{A}(x),\mathcal{B}(x), \mathcal{C}(x)$ from our solution (\ref{soln}), it turns out that both of the conditions ensuring absence of conical singularities are automatically satisfied (for all $a,b,c,d, \gamma^{(1)}_{x\psi}$!). Therefore, we have shown that our deformation $\gamma^{(1)}_{ab}$ extends to a smooth tensor on $S^3$, as required.

Let us now compute the trace of $\gamma^{(1)}_{ab}$ of our solution. We find
\be
\gamma^{(1)} =\tfrac{1}{4} k^2  \left[ 2d (1+x^2)+ 4c + 3ax (1-2x^2) +b x(-3+2x^2) - 2((1-x^2)\gamma^{(1)}_{x\psi})' \right]  \; .
\ee 
Integrating this over $S$ gives,
\be
\Theta = \frac{128 \pi^2 (d+3c)}{3k}   \; .
\ee
Therefore, for $S$ to be a MTS we need
\be
d+3c>0 \; .
\ee
Observe that $\gamma^{(1)}_{x\psi}$ cancels out in $\Theta$ (this follows from smoothness of  $\gamma^{(1)}_{x\psi}$ at $x=\pm 1$). This  must be the case since $\Theta$ is gauge invariant, while $\gamma^{(1)}_{x\psi}$ is not. Curiously, the parameters $a,b$ also do not appear in $\Theta$.   Of course, it is only the sign of the parameter $d+3c$ which is physical and its precise value can be changed by an overall scaling of our solution.   Thus, we really have a three parameter family of smooth deformations of the extreme MP horizons.  

To summarise, we have shown that the general smooth deformation that satisfies (\ref{g1abeqn}) for the extreme MP horizon with equal angular momenta, is a three parameters family specified by $a,b,c,d$ with the scale is set by $d+3c$.  Thus we find a three-dimensional moduli space of solutions, in line with our general result.
The $a=b=d=0$ solution corresponds to the $SU(2)\times U(1)$ invariant deformations arising from the known black holes discussed in section (\ref{sec:homodef}). We are not aware of any known black hole solutions which correspond to our more general deformations. It is natural to ask if this indicates the existence of new black hole solutions. We will discuss this below.

\section{Discussion}

The question that motivated this work was ``what is the set of extreme black holes with a prescribed near-horizon geometry?"  In this paper, we have shown that we may gain an essentially complete understanding of the space of possible {\it infinitesimal} transverse deformations of a near-horizon geometry.  However, to address the original question one needs to (a) understand the space of {\it exact} transverse deformations and (b) determine which extend out to asymptotically flat (or KK, AdS...) spacetimes. Since we have only solved (a) infinitesimally, we are not able to address (b) in this work. We will return to these questions at the end.

Perhaps our most interesting finding is the three-parameter family of transverse deformations of the extreme Myers-Perry horizon with equal angular momenta (\ref{homohor}) found in section (\ref{sec:toricdef}). These do not correspond to any known black hole solution. In view of the above comments we are of course unable to determine whether they correspond to new black holes or not. However, it is interesting to ask what black holes these would correspond to. By construction it is the most general deformation which preserves $U(1)^2$ rotational symmetry of the near-horizon geometry. Therefore the spacetime symmetry is $\mathbb{R}\times U(1)^2$. As discussed in the introduction, such spacetimes are understood up to the issue of determining what rod structures are actually realised by regular black holes.  

By the uniqueness theorem~\cite{Hollands:2007aj}, asymptotically flat black holes with the same rod structure as the Myers-Perry black hole must be isometric to the Myers-Perry black hole.  Therefore, we deduce that our deformations would correspond to black holes with a distinct rod structure.  In fact, the most general rod structure compatible with asymptotic flatness and an $S^3$ horizon corresponds to a spacetime with a DOC containing an arbitrary number of 2-cycles. Therefore, these black holes would correspond to new vacuum black holes with bubbles in the DOC. Exact solutions of this form are not known for the vacuum equations, although supersymmetric solutions of this kind are known to exist~\cite{Kunduri:2014iga}.  

There is another possible interpretation of these solutions. By taking a suitable quotient by a subgroup  $\mathbb{Z}_p$ of the $U(1)^2$-action, the horizon geometry (\ref{homohor}) can have lens space topology $L(p,q)$. Since this corresponds  to a symmetry of the solution, our deformation also gives smooth deformations of such horizons.  In fact, it is an open problem to determine the existence of asymptotically flat vacuum black holes with lens space topology. However, if extreme ones do exist, their near-horizon geometries would be locally isometric to either the Myers-Perry black holes or the KK black holes~\cite{Kunduri:2008rs}.  If there exists a black lens with near-horizon geometry locally isometric to the homogeneous geometry (\ref{homohor}), then our deformations would capture such a solution. Again, we note that exact supersymmetric solutions of this kind are known~\cite{Kunduri:2014kja}.  

There are a number of natural extensions of the above work. Most obviously, classifying the $U(1)^2$-invariant deformations of the remaining 5d vacuum near-horizon geometries (i.e. Myers-Perry with unequal angular momenta, the fast rotating KK black hole and the Kerr string/black ring~\cite{Kunduri:2008rs}). Given the present work, it seems likely all these cases can be solved explicitly. Furthermore, it would be interesting to investigate what information pertaining to the rod structure is retained in our infinitesimal deformations. This would then allow one to test the above possible interpretations. 

More generally, it is clear that in principle our method could be applied to solutions of Einstein's equations with an energy-momentum tensor.  For example, it seems plausible that a complete understanding of (axisymmetric) transverse deformations of extreme horizons in 4d Einstein-Maxwell theory is achievable. This may be interesting, since presumably the local uniqueness theorem we found for vacuum solutions would not extend to this case due to the possibility of black holes immersed in background fields (e.g. the Kerr-Newman-Melvin solution). A natural conjecture would be that the most general axisymmetric deformation of an extreme Kerr-Newman horizon would correspond to that of the extreme Kerr-Newman-Melvin solution (which includes Kerr-Newman)\footnote{The near-horizon geometry of the extreme Kerr-Newman-Melvin black hole must be isometric to that of the Kerr-Newman black hole by the near-horizon uniqueness theorems. This has also been confirmed explicitly~\cite{Booth:2015nwa}.}.  

It would also be interesting to study deformations of extreme horizons in supergravity theories.  It seems plausible that progress could be made for supersymmetric deformations of supersymmetric near-horizon geometries in five dimensional minimal supergravity~\cite{Reall:2002bh}.

It is of course also of interest to apply our method to solutions with a cosmological constant. We emphasise that the linearised Einstein equation (\ref{g1abeqn}) is also valid in the presence of a cosmological constant. For AdS solutions this could help one investigate the possibility of black holes with a single Killing field corotating with the horizon conjectured to exist in~\cite{Kunduri:2006qa}.\footnote{Recently, examples of such solutions in four dimensions have been constructed numerically~\cite{Dias:2015rxy}.} For example, one could study non-axisymmetric transverse deformations of the extreme Kerr-AdS horizon. This would be a PDE problem though, so more complicated than the examples studied in this paper. 

Finally, let us return to the question (a) posed at the start of this section. Recall we have studied transverse deformations of a near-horizon geometry (\ref{spacetime}) to first order in the scaling parameter $\varepsilon$. A natural strategy in line with the approach we have taken would to be to examine the second and higher order transverse deformations, $g^{(n)} = \frac{d^n g_{\varepsilon}}{d\varepsilon ^n} |_{\varepsilon=0}$ for $n \geq 2$. By taking the corresponding derivatives of the Einstein equation $\text{Ric}(g_{\varepsilon})=\Lambda g_{\varepsilon}$, one obtains a linear elliptic PDE at each order for $g^{(n)}$ which depends on the lower order data $\bar{g}, g^{(1)}, \dots, g^{(n-1)}$.  Now, even if one can find a solution at each order, one would then have to study the convergence properties of the series in $\varepsilon$. Thus ultimately this method may be better suited for revealing obstructions to transverse deformations. For instance, it seems plausible that for the toroidal horizon in section (\ref{sec:toroidal}), one could construct an inductive argument along the lines of~\cite{Moncrief:1983xua}\footnote{Their result assumes the null generators are periodic so does not immediately apply here.}, to show that any  spacetime with such a near-horizon geometry is a plane wave solution (\ref{planewave}) and hence not a black hole.
\\

\noindent {\bf Acknowledgements}. We would like to thank Jos\'e Figueroa-O'Farrill, Hari Kunduri and Simon Ross for useful comments. CL was supported by a Principal Career Development Scholarship at the University of Edinburgh. JL is supported by the Science and Technology Facilities Council (STFC) [ST/L000458/1]. 

\appendix

\section{Linearised Einstein equations}

The components of the linearised Ricci tensor $R^{(1)}_{\mu\nu}$ defined by (\ref{Ric1}), for the perturbation $g^{(1)}$ (\ref{g1}) in the background near-horizon geometry $\bar{g}$ (\ref{nhg}), in our basis (\ref{basis}), are:
\bea
R^{(1)}_{--} &=& 0 \label{Rmm} \\
R^{(1)}_{-a} &=& h^{(1)}_a - \frac{1}{2} h^{b} \gamma^{(1)}_{ab} + \frac{1}{2} {D}^b \gamma^{(1)}_{ab} - \frac{1}{2} {D}_a  \gamma^{(1)} + \frac{1}{4} h^{}_a  \gamma^{(1)}   \label{Rma} \\ 
\nonumber R^{(1)}_{+-} &=&  r \left[ 3 F^{(1)} - 3 h^{a} h_a^{(1)} + {D}^a h_a^{(1)} + h^{a} h^{b} \gamma^{(1)}_{ab}  - \frac{1}{2} h^{(a} {D}^{b)} \gamma^{(1)}_{ab} \right. \\
& & \left. - \frac{1}{2}\left( {D}^{(a} h^{b)} \right) \gamma^{(1)}_{ab} + \frac{1}{2} F  \gamma^{(1)}  + \frac{1}{4} h^{a} \left( {D}_a  \gamma^{(1)}   - h^{}_a  \gamma^{(1)}  \right)\right] \label{Rpm} \\
\nonumber R^{(1)}_{ab} &=& r \left[-4 h^{}_{(a} h^{(1)}_{b)} + 2 {D}_{(a} h^{(1)}_{b)} + F \gamma^{(1)}_{ab} - {h^{}}^2 \gamma^{(1)}_{ab} + 2 h^{}_{(a} h^{c} \gamma^{(1)}_{b)c} \right. \\
\nonumber & & + \frac{1}{2} {D}^c h^{}_c \gamma^{(1)}_{ab}  - \left( {D}^c h^{}_{(a}\right) \gamma^{(1)}_{b)c} + \frac{3}{2} h^{c}  {D}_c \gamma^{(1)}_{ab} - 2 h^{c} {D}_{(a}  \gamma^{(1)}_{b)c} \\
\nonumber & & - h^{}_{(a} {D}^c  \gamma^{(1)}_{b)c} - \frac{1}{2} {D}^2 \gamma^{(1)}_{ab}  + {D}_{(a}  {D}^c  \gamma^{(1)}_{b)c} - \frac{1}{2} h^{}_a h^{}_b \gamma^{(1)} + \frac{1}{2} \left( {D}_{(a} h^{}_{b)} \right) \gamma^{(1)} \\
& &   \left.  + h^{}_{(a} {D}_{b)}\gamma^{(1)} - \frac{1}{2} {D}_{(a} {D}_{b)}\gamma^{(1)} + {R}_{(a}^{~ ~ c} \gamma^{(1)}_{b)c} - {R}_{a~b~}^{~c~d}\gamma^{(1)}_{cd} \right] \label{Rab} 
\eea
\bea
\nonumber R^{(1)}_{+a} &=& r^2 \left[ -3 h^{}_a F^{(1)} + \frac{3}{2} {D}_a F^{(1)} - 2 {h^{}}^2 h^{(1)}_a + \frac{1}{2} F h^{(1)}_a  + {D}^b h^{}_b h^{(1)}_a \right. \\
\nonumber & & + \left( {D}^b  h^{}_a    \right) h^{(1)}_b  + \frac{3}{2} h^{}_a  h^{b} h^{(1)}_b + \frac{1}{2} \hat{R}_{a}^{~b}  h^{(1)}_b + 2  h^{b} {D}_b  h^{(1)}_a   \\
\nonumber & &  - \frac{3}{2} {D}_a \left(  h^{b}  h^{(1)}_b \right)  - \frac{1}{2}  h^{}_a  {D}^b h^{(1)}_b  + \frac{1}{2} {D}_a {D}^b  h^{(1)}_b    - \frac{1}{2} {D}^2  h^{(1)}_a \\
\nonumber & &  - \frac{1}{2} {D}^b F \gamma^{(1)}_{ab} + \frac{3}{4} F h^{b}   \gamma^{(1)}_{ab}   +  h^{}_c  {D}^{[b}  h^{c]} \gamma^{(1)}_{ab} + 3  h^{}_b  {D}_{[a}  h^{}_{c]} \gamma^{(1)bc} \\
\nonumber & &   -   {D}_{b} {D}_{[a}  h^{}_{c]} \gamma^{(1)bc} - \frac{1}{4} F  {D}^{b} \gamma^{(1)}_{ab} +  {D}^{[b}  h^{c]} {D}_{b} \gamma^{(1)}_{ac}  +  {D}_{[b}  h^{}_{a]} {D}_{c} \gamma^{(1)bc} \\
\nonumber & & + \frac{1}{4} \left( {D}_{a} F \right) \gamma^{(1)} - \frac{3}{8} F  h^{}_{a} \gamma^{(1)}  + \frac{1}{2}  h^{b} {D}_{[b}  h^{}_{a] }  \gamma^{(1)}  \\
& & \left. + \frac{1}{4} F {D}_a  \gamma^{(1)}  - \frac{1}{2}  {D}_{[b}  h^{}_{a]}  {D}^b \gamma^{(1)} \right] \label{Rpa} 
\eea
\bea
\nonumber R^{(1)}_{++} &=& r^3 \left[ F F^{(1)} - \frac{7}{2} {h^{}}^2 F^{(1)} + \frac{3}{2}  {D}^a  h^{}_a F^{(1)} + \frac{5}{2}  h^{a} {D}_a F^{(1)} - \frac{1}{2} {D}^2 F^{(1)} \right. \\
\nonumber & & + {D}^a F h^{(1)}_a - h^{a}  F h^{(1)}_a - 2 h^{}_a  {D}^{[a}  h^{b] } h^{(1)}_b + 2 {D}^{[a}  h^{b]} {D}_{[a}  h^{(1)}_{b]} \\
\nonumber & & + \frac{3}{2} F h^{a} h^{b} \gamma^{(1)}_{ab} - 2 h^{(a} \left( {D}^{b)} F  \right)\gamma^{(1)}_{ab} - 2 {D}^{[a}  h^{b]} {D}_{[a}  h^{c]} \gamma^{(1)}_{bc} \\
\nonumber & & + \frac{1}{2}  {D}^{(a} F  {D}^{b)} \gamma^{(1)}_{ab}  - \frac{1}{2} h^{(a} F  {D}^{b)} \gamma^{(1)}_{ab} + \frac{1}{2} {D}^{a} {D}^{b} F  \gamma^{(1)}_{ab}  \\
\nonumber & & - \frac{1}{2} F   {D}^{(a}  h^{b)} \gamma^{(1)}_{ab} + \frac{1}{4} \left( - {D}^{a} F {D}_{a} \gamma^{(1)} + {D}^{a} F h^{}_a \gamma^{(1)} \right. \\
& & \left. \left. + h^{a}  F  {D}_{a} \gamma^{(1)} - {h^{}}^2 F   \gamma^{(1)} \right) \right]  \label{Rpp}
\eea
where we have defined the trace $\gamma^{(1)} = \gamma^{ab} \gamma^{(1)}_{ab}$.

The resulting $-a, +-, ab$ components of the vacuum Einstein equations are given in simplified form in (\ref{h1}), (\ref{f1eqn}) and (\ref{g1abeqn}). The $+a$ and $++$ components \eqref{Rpa} and \eqref{Rpp} of the linearised Einstein equation are in fact redundant due to the linearised contracted Bianchi identity, as we now show.

The contracted Bianchi identity for a 1-parameter family of spacetimes $(M,g(\varepsilon))$ is given by
\be
0 = \nabla^\nu G_{\nu \rho}= g^{\mu \nu} \nabla_\mu \left(R_{\nu \rho} - \frac{1}{2} g_{\nu \rho} R \right) 
\ee
where for notational simplicity we suppress the explicit $\varepsilon$ so $\nabla, G_{\mu\nu}, R_{\mu\nu}, R$ are the metric connection, Einstein tensor, Ricci tensor and Ricci scalar of $g(\varepsilon)$.  We will be concerned with solutions to the Einstein equation $R_{\mu\nu} = \Lambda g_{\mu\nu}$. For convenience we define the modified Einstein tensor $E_{\mu\nu} \equiv G_{\mu\nu} - \Lambda (1- \tfrac{1}{2} D) g_{\mu\nu}$, so that the Einstein equation is equivalent to $E_{\mu\nu}=0$ and the contracted Bianchi identity is $\nabla^\mu E_{\mu \nu}=0$. (Of course this step is unnecessary for vacuum solutions $\Lambda=0$).

The linearised contracted Bianchi identity is thus
\be
0 = \left. \frac{\td (\nabla^\mu E_{\mu \rho}) }{\td \varepsilon} \right|_{\varepsilon=0} \; .  \label{lincbidef}
\ee
The connection $\nabla$ of $g(\varepsilon)$ must be related to the connection $\bar{\nabla}$ of $\bar{g} \equiv  g(0)$ by a tensor $C^\rho_{\mu \nu} (\varepsilon)$ such that for say a covector $\omega_\rho$, 
\be
\nabla_\mu \omega_\nu = \bar{\nabla}_\mu \omega_\nu - C^\rho_{\mu \nu} (\varepsilon) \omega_\rho \; ,
\ee
where $C^\rho_{\mu \nu} (0) =0$. Equation \eqref{lincbidef} thus expands into 
\be
0 = \left. \frac{\td }{\td \varepsilon} \left[  g^{\mu \nu } \left( \bar{\nabla}_\mu E_{\nu \rho}  - C^\sigma_{\mu \nu} (\varepsilon) E_{\rho \sigma} -  C^\sigma_{\mu \rho} (\varepsilon) E_{ \sigma \nu} \right) \right]  \right|_{\varepsilon=0} \; , 
\ee
which becomes
\be
0 =\left. \frac{\td g^{\mu \nu}}{\td \varepsilon} \right|_{\varepsilon=0} \bar{\nabla}_\mu \bar{E}_{\nu \rho} + \bar{g}^{\mu \nu} \left( \bar{\nabla}_\mu \left. \frac{\td E_{\nu \rho}}{\td \varepsilon} \right|_{\varepsilon=0} - \left. \frac{\td C^\sigma_{\mu \nu}}{\td \varepsilon} \right|_{\varepsilon=0} \bar{E}_{\rho \sigma} - \left. \frac{\td C^\sigma_{\mu \rho}}{\td \varepsilon} \right|_{\varepsilon=0} \bar{E}_{ \sigma \nu} 
\right)  \; ,
\ee 
where $\bar{E}$ is the modified Einstein tensor of $\bar{g}$.  For Einstein solutions $\bar{E}_{\mu\nu}=0$ this simplifies to just
\be
\bar{\nabla}^\mu E^{(1)}_{\mu\nu}=0
\ee
where we have defined $ E^{(1)}_{\mu\nu} =  \left. \frac{\td E_{\mu \nu}}{\td \varepsilon} \right|_{\varepsilon=0}$. To evaluate the linearised Einstein tensor we need,
\be
\left. \frac{\td R}{\td \varepsilon} \right|_{\varepsilon=0} =  \left. \frac{\td (g^{\rho \sigma} R_{\rho \sigma})}{\td \varepsilon} \right|_{\varepsilon=0}= \left. \frac{\td g^{\rho \sigma}}{\td \varepsilon} \right|_{\varepsilon=0} \bar{R}_{\rho \sigma} + \bar{g}^{\rho \sigma} R^{(1)}_{\rho \sigma}  =  - \Lambda \bar{g}^{\rho \sigma} g^{(1)}_{\rho \sigma}+ \bar{g}^{\rho \sigma} R^{(1)}_{\rho \sigma} 
\ee
where we have used the definition (\ref{linE}) and  the Einstein equation $\bar{R}_{\mu\nu}=\Lambda \bar{g}_{\mu\nu}$. The linearised modified Einstein tensor is therefore,
\bea
E^{(1)}_{\mu\nu} &=& R^{(1)}_{\mu\nu} - \tfrac{1}{2} \bar{g}_{\mu\nu} \left. \frac{\td R}{\td \varepsilon} \right|_{\varepsilon=0}- \tfrac{1}{2} g^{(1)}_{\mu\nu} \bar{R}  - \Lambda (1-\tfrac{1}{2}D)g^{(1)}_{\mu\nu} \nn \\
&=& R^{(1)}_{\mu\nu}- \Lambda g^{(1)}_{\mu\nu} - \tfrac{1}{2} \bar{g}_{\mu\nu} \bar{g}^{\rho \sigma}( R^{(1)}_{\rho \sigma}-\Lambda g_{\rho \sigma})  \; ,
\eea
where again we have made use of the Einstein equations for $\bar{g}$.  In summary, we have shown that the linearised contracted Bianchi identity for an Einstein background is
\be
0= \bar{\nabla}^{\mu} (R^{(1)}_{\mu \nu}-\Lambda g^{(1)}_{\mu\nu}) - \tfrac{1}{2} \bar{\nabla}_\nu ( \bar{g}^{\alpha \beta} (R^{(1)}_{\alpha \beta}-\Lambda g^{(1)}_{\alpha \beta})) \; . \label{linbi} 
\ee
Observe this derivation was valid for any 1-parameter family spacetimes such that $g(0)$ is obeys the Einstein equation.

We may now evaluate this for our 1-parameter family spacetimes (\ref{spacetime}).  Let us write the components of  $R^{(1)}_{\mu \nu}- \Lambda g^{(1)}_{\mu \nu}$ as,
\bea
&&R^{(1)}_{++} - \Lambda g^{(1)}_{++} = r^3 S_{++} \;, \qquad R^{(1)}_{+a} - \Lambda g^{(1)}_{+a} = r^2 S_{+a} \; ,\qquad R^{(1)}_{ab}- \Lambda g^{(1)}_{ab} = r S_{ab} \; , \nn \\ &&R^{(1)}_{+-} = r S_{+-} \; , \qquad R^{(1)}_{-a} = S_{-a}   \; ,
\eea
where $S_{\mu \nu}$ is independent of $r$. The $+,-,a$ components of \eqref{linbi} are,  
\bea
3 S_{++} & =&   3 h^{a} S_{+a} - D^a S_{+a} - \frac{1}{2} (D^a F + h^{a} F )S_{-a} - \frac{1}{4} F \gamma^{ab} S_{ab} \; , \label{linbip}\\
 S^a_{~a} &=&  2 D^a S_{-a} - 4 h^{a}  S_{-a}  \; ,\label{linbim}\\
\nn S_{+a} &=& \frac{1}{2} F  S_{-a} + D_{[a} h_{b]} S^{~b}_{-} - \frac{1}{2}  h_a S_{+-} + \frac{1}{2} D_a S_{+-} - \frac{1}{2 }h_{a} S_{+-} \\
& & + h^{b} S_{ab} - \frac{1}{2} D^b S_{ab} + \frac{1}{4} D_a (  \gamma^{bc} S_{bc} ) - \frac{1}{4} h_a \gamma^{bc} S_{bc}   \; ,
\label{linbia}
\eea
respectively. Equations \eqref{linbip} and \eqref{linbia} thus state that the $+a$ and $++$ components of the linearised Einstein equation (\ref{linE}) can be expressed entirely in terms of the other non-trivial components of the linearised Einstein equation (this can also be verified directly using \eqref{Rma}-\eqref{Rpp}). Furthermore, equation \eqref{linbim} explains the tracelessness of the linearised Einstein equation \eqref{g1abeqn}. 

\section{Three-dimensional spacetimes}

In this section we show that our first order analysis applied in three spacetime dimensions agrees with the exact analysis obtained in~\cite{Li:2013pra}.

In three dimensions, the cross-section $S$ are 1-dimensional and hence have no curvature. Thus we can write the horizon data as
\be
\gamma = \td x^2, \qquad h = h(x) \td x
\ee
and the horizon equations (\ref{horeq}) and (\ref{F}) simplify to 
\be
h'(x) - \tfrac{1}{2} h(x)^2=\Lambda \; ,\qquad F(x) = \tfrac{1}{2}h(x)^2 - \tfrac{1}{2} h'(x) +\Lambda \; . 
\ee
Since we are in 1-dimension all tensors are equivalent to scalars, so we will not make the distinction.

As noted in the main text the linearised Einstein equation (\ref{g1abeqn}) is trivially satisfied for $D=3$, so $\gamma^{(1)}(x)$ may be any function of $x$. The rest of the first order data is then determined by (\ref{h1}) and (\ref{f1eqn}), which simplifies to
\be
h^{(1)} = \tfrac{1}{4} h \gamma^{(1)} , \qquad F^{(1)}  =  0 
\ee
where we have used the horizon equations.

The exact general solution in Gaussian null coordinates (\ref{gnc}) was obtained in~\cite{Li:2013pra} and is given by,
\bea
\gamma (r,x)  &=& (1+r \gamma_1(x))^2  \\
h(r, x)  &=& h(x) (1+\tfrac{1}{2} r \gamma_1(x)) \\
F(r,x)  &=& \tfrac{1}{4} h(x)^2 +\tfrac{1}{2} \Lambda
\eea
where $\gamma_1(x)$ is an arbitrary function. It is easy to see this leads to the first order data above with $\gamma^{(1)}(x) = 2 \gamma_1(x)$, as required.

It is worth remarking that the condition for $S$ to be a MTS is $\gamma^{(1)}>0$. This coincides with the assumption $\gamma_1(x)>0$ made in that work.  Thus, the results of~\cite{Li:2013pra}  show that the most general negative Einstein spacetime containing a non-singular degenerate Killing horizon such that $S$ is a MTS, is globally isometric to the extreme BTZ black hole.

\section{Extreme Kerr black hole in Gaussian null coordinates}

Consider the extreme Kerr black hole in BL coordinates $(t,r,\theta,\phi)$,
\be
ds^2= -\frac{\Delta}{\rho^2} (\td t - a \sin^2 \theta \td\phi)^2  +\frac{\sin^2 \theta}{\rho^2} [a \td t - (r^2+a^2) \td\phi]^2 + \frac{\rho^2 \td r^2}{\Delta} + \rho^2 \td\theta^2 \; ,
\ee
with $\rho^2=r^2 +a^2 \cos^2\theta$ and $\Delta = (r-a)^2$, where $a>0$ is the rotation parameter. The surface $r=0$ is a smooth degenerate horizon.
 The Killing vector null on the horizon is
\be
V= \frac{\partial}{\partial t}+ \Omega_H \frac{\partial}{\partial \phi}  
\ee
where $\Omega_H = 1/(2a)$ is the angular velocity of the hole relative to infinity.   

We will construct Gaussian null coordinates by explicitly constructing null geodesics that shoot out of the horizon. Thus, consider geodesic curves $\gamma(\lambda)=(t(\lambda), r(\lambda), \theta(\lambda), \phi(\lambda))$ in this geometry where $\lambda$ is an affine parameter. Because $T=\partial_t$ and $m=\partial_\phi$ are Killing vectors the following are constants along geodesics
\bea
E &=& T \cdot \dot{\gamma} = -\frac{\Delta}{\rho^2} (\dot{t}-a\sin^2\theta \dot{\phi})+ \frac{a \sin^2 \theta}{\rho^2}[ a\dot{t}- (r^2+a^2) \dot{\phi}] \\
J &=& m \cdot \dot{\gamma} = \frac{a \Delta \sin^2\theta}{\rho^2} (\dot{t} -a \sin^2 \theta \dot{\phi}) - \frac{(r^2+a^2)\sin^2\theta}{\rho^2} [a \dot{t}-(r^2+a^2) \dot{\phi}]  \;.
\eea
We wish to find null geodesics such that $V \cdot \dot{\gamma}=1$ and $m \cdot \dot{\gamma}=0$. The latter condition is required since we wish to find coordinates adapted to an axisymmetric cross-section $S$ so $m$ must be tangent to $S$. These conditions correspond to $E=1$ and $J=0$. The above can then be solved  for $\dot{t}$ and $\dot{\phi}$,
\bea
\label{tdot}
\dot{t} &=& - \frac{ (r^2+a^2)^2 - \Delta a^2 \sin^2\theta}{\rho^2 \Delta} \\
\dot{\phi} &=&-\frac{2ra^2}{\rho^2 \Delta}  \; . \label{phidot}
\eea
Now inserting these into the null constraint and simplifying gives
\be
\label{null}
\dot{r}^2+ \Delta \dot{\theta}^2 = \frac{(r^2+a^2)^2 - \Delta a^2 \sin^2\theta}{\rho^4}  \; .
\ee
We need one other coupled ODE for $r, \theta$ to uniquely determine the geodesics. 
For instance, the $r$-equation gives,
\be
\label{req}
\ddot{r}= \frac{2 a^2 \sin \theta \cos \theta \; \dot{r}\dot{\theta}}{\rho^2}  +\left( \frac{2r \Delta}{\rho^2} - \frac{\Delta'}{2} \right) \dot{\theta}^2 - \frac{a^2\Delta' \sin^2 \theta}{2\rho^4} - \frac{4a^3 r^2 \sin^2 \theta}{\rho^6}  \; ,
\ee
which we have simplified by eliminating the $\dot{r}^2$ terms using (\ref{null}).

We wish to solve the coupled ODEs (\ref{null}) and (\ref{req}) subject to the following boundary conditions:
\bea
r(0)=a \qquad \theta(0)=\Theta  \qquad \dot{\theta}(0)= 0  \;.
\eea
This choice is required to ensure that the $\lambda \to 0$ limit of the $\theta$ coordinate coincides with a new coordinate defined on the horizon which we have called $\Theta$. 
For every initial value $\Theta$, this system yields a unique solution $r(\lambda, \Theta)$ and $\theta(\lambda, \Theta)$. Then, integrating (\ref{tdot}) and (\ref{phidot}) wrt $\lambda$ gives,
\bea
t &=& v+ f(\Theta) - \int  \frac{ (r^2+a^2)^2 - \Delta a^2 \sin^2\theta}{\rho^2 \Delta}  d\lambda  \; , \\
\phi &=& \chi +\Omega_H v +g(\Theta)- \int \frac{2ra^2}{\rho^2 \Delta} d\lambda  \; ,
\eea
where the integrands are treated as functions of $(\lambda, \Theta)$ and $v, \chi, f(\Theta), g(\Theta)$ are integration constants and functions chosen so that
\be
V= \frac{\partial}{\partial v} \; ,  \qquad m = \frac{\partial}{\partial \chi} \; .  \label{killing}
\ee
The former is required by the definition of Gaussian null coordinates, whereas the latter we have chosen so that the coordinates are adapted to the rotational Killing field $m$.  

The final condition which must be imposed is that the geodesics $\dot{\gamma}$ are orthogonal to a cross-section $v=0, \lambda=0$. Such a cross-section has coordinates $(\Theta, \chi)$ so this requires 
\be
g_{\lambda \Theta}=0  \label{orth} \; .
\ee
Recall we have already imposed $g_{\lambda \chi} = m \cdot \dot{\gamma} =0$, so there are no further conditions. Condition (\ref{orth}) is a complicated condition which requires knowledge of the solution $r(\lambda, \Theta), \theta(\lambda, \Theta)$ to the geodesic equations. The coordinates $(v,\lambda, \Theta, \chi)$ are our required Gaussian null coordinates and $(\Theta, \chi)$ are coordinates on a cross-section.

In practice it is easiest to find this coordinate system by expanding in a power series in $\lambda$. We seek an expansion of the form
\bea
r(\lambda)=a+ \sum_{n=1}^\infty \frac{a_n \lambda^n}{n!} \; , \qquad \qquad \theta(\lambda) = \Theta+\sum_{n=2}^\infty \frac{b_n \lambda^n}{n!}  \; ,
\eea
where we have implemented the initial conditions. The $O(1)$ term in equation (\ref{null}) gives,
\be
a_1^2 = \frac{4}{(1+\cos^2 \Theta)^2}
\ee
whereas the $O(1)$ term in (\ref{req}) (or $O(\lambda)$ term in (\ref{null})) gives
\be
a_2=- \frac{2\sin^2 \Theta}{a(1+\cos^2 \Theta)^3}  \; .
\ee
Hence, choosing the appropriate sign for $a_1$, so far we have shown,
\bea
r(\lambda, \Theta) &=& a + \frac{2}{1+\cos^2 \Theta} \lambda - \frac{\sin^2 \Theta}{a(1+\cos^2 \Theta)^3} \lambda^2 
+ O(\lambda^3)  \\
\theta(\lambda, \Theta) &=&\Theta +O(\lambda^2)  \; .
\eea
It turns out these are the only terms we will need.
We may now integrate to find $t,\phi$,
\bea
t(v,\lambda, \Theta)&=& v +f(\Theta) + \frac{a^2 (1+\cos^2 \Theta)}{\lambda} - 2a \log \lambda + O(\lambda) \; ,\\
\phi(v, \chi, \lambda, \Theta) &=&  \chi+ \frac{v}{2a}+ g(\Theta) + \frac{a(1+\cos^2\Theta)}{2\lambda} +O(\lambda)  \; .
\eea
We may now implement the condition (\ref{orth}). Since by construction $\partial_\lambda$ is null and geodesic, it is sufficient to impose this at $\lambda=0$. The condition for this turns out to be,
\be
f'(\Theta) = - \frac{4a \sin \Theta \cos^2\Theta}{(1+\cos^2\Theta)^2}
\ee
which fixes $f(\Theta)$ up to an irrelevant integration constant.  With this choice our coordinate system is of Gaussian null type.  However, we have not fully fixed the coordinates on the horizon. From (\ref{killing}) we already have $g_{\chi\chi}= g_{\phi \phi}$ (for all $\lambda$ in fact). To fully fix the coordinates on the horizon  we impose the condition $g_{\Theta \chi}=0$ at $\lambda=0$, which turns out to be,
\be
g'(\Theta) = \frac{ 2 \sin\Theta \cos \Theta}{(1+\cos^2\Theta)^2}  \; ,
\ee
which fixes $g(\Theta)$ up to an irrelevant integration constant.
It is then readily verified that $g_{\Theta\Theta}= g_{\theta\theta} $ at $\lambda=0$, as required. In summary, the explicit metric on a cross-section $v=0, \lambda=0$ is,
\be
\gamma_{ab} dx^a dx^b = a^2 (1+\cos^2\Theta) \td\Theta^2 + \frac{ 4 a^2 \sin^2\Theta}{(1+\cos^2\Theta)^2}\td\chi^2  \; .
\ee
Upon the coordinate change $x= \cos\Theta$, this agrees with the horizon metric used in the main text (\ref{kerrhor}).

We are now ready to compute the first order data $\gamma^{(1)}_{ab}= \partial_\lambda \gamma_{ab} |_{\lambda=0}$, which requires $\gamma_{ab}(\lambda, x)$ for small $\lambda$. Since $\gamma_{ab} = g_{ab}$, we need the $\Theta \Theta, \Theta \chi, \chi\chi$ components of the spacetime metric to first order in $\lambda$. 
Using the above coordinate change we find,
\be
\gamma^{(1)}_{\Theta\Theta} = \frac{4a}{1+\cos^2\Theta}, \qquad \gamma^{(1)}_{\Theta \chi} = - \frac{4 a\sin^3\Theta \cos\Theta}{(1+\cos^2\Theta)^2}, \qquad \gamma^{(1)}_{\chi \chi}= \frac{16 a \sin^2 \Theta \cos^2\Theta}{(1+\cos^2\Theta)^3}
\ee
Using the coordinate change $x= \cos\Theta$ this gives (\ref{kerrgam1}).

\section{5D black holes in Gaussian null coordinates}

In this Appendix we list all known vacuum extreme black holes with a homogeneous near-horizon geometry (\ref{homohor}).

\subsection{Extreme Myers-Perry}
The five-dimensional extreme Myers-Perry black hole with equal angular momenta is a one-parameter family of solutions,
\be
ds^2 = - f^2 \td t^2 + g^2 \td r^2 + \tfrac{1}{4} h^2 (\sigma_3 - \Omega \td t)^2 + \tfrac{1}{4} r^2 (\td\theta^2 +\sin^2\theta \td\phi^2)
\ee
where  $f^2= r^2/(h^2 g^2)$,
\be
g^{-2} = \frac{(r^2-2a^2)^2}{r^4} , \qquad h^2 = r^2 \left(1+ \frac{4a^4}{r^4} \right), \qquad \Omega= \frac{8 a^3}{r^2 h^2} \; ,
\ee
where  $\sigma_3 = \td\psi+ \cos \theta \td\phi$ and $a$ is a non-zero parameter. The surface $r= \sqrt{2} a$ is a smooth degenerate horizon. Let us convert to coordinates which are regular on this horizon. 

The Killing field which is null on the horizon is,
\be
V = \partial_t + \Omega_H \partial_\psi  \; ,
\ee
where $\Omega_H = \Omega|_{r= \sqrt{2} a} = a^{-1}$.  Now change coordinates $(t,\psi)$ to $(v,\psi')$ defined by
\be
\td v = \td t + \frac{g}{f} \td r, \qquad \td\psi' = \td\psi - \Omega_H \td t -  (\Omega- \Omega_H) \frac{g}{f} \td r   \; .
\ee
We get
\be
ds^2 = -f^2 \td v^2 + \frac{2r}{h} \td v \td r+ \tfrac{1}{4} h^2 (\sigma'_3 - (\Omega-\Omega_H) \td v)^2 + \tfrac{1}{4} r^2 (\td\theta^2 +\sin^2\theta \td\phi^2)  \; .
\ee
The metric in these  coordinates is regular on the future horizon.   It is now clear that the metric can be now put into Gaussian null coordinates by simply defining a new radial coordinate,
\be
\lambda = \int^r_{\sqrt{2} a} \frac{r'}{h(r')} \td r'  \; .
\ee
This coordinate change is well defined everywhere outside and on the horizon since $\td\lambda /\td r= r/h>0$. Inverting this we may write $r= r(\lambda)$. For instance, near the horizon $r = \sqrt{2}( a +\lambda+ O(\lambda^2))$.  

The coordinates $(v,\lambda, \theta, \phi, \psi')$ are Gaussian null coordinates and $x^a=( \theta, \phi, \psi')$ are coordinates on a cross-section of the horizon. The metric is given by (\ref{gnc}) (with $r$ replaced by  $\lambda$ of course) and the data,
\bea
F(\lambda, x) &=& \frac{-f^2+ \tfrac{1}{4} h^2 (\Omega- \Omega_H)^2}{\lambda^2}, \qquad h_a(\lambda, x)dx^a = - \tfrac{1}{4}\frac{h^2 (\Omega- \Omega_H)}{\lambda} \sigma_3' \nn \\
\gamma_{ab}(\lambda, x) \td x^a \td x^b &=& \tfrac{1}{4} h^2 \sigma_3'^2+ \tfrac{1}{4} r^2 (\td\theta^2 +\sin^2\theta \td\phi^2)  \; .
\eea
It is easy to see that all these quantities are smooth at the horizon $\lambda=0$.

The near-horizon geometry is easily extracted by evaluating the data at $\lambda=0$. For instance, the horizon metric is
\be
\gamma_{ab}(x)\td x^a \td x^b = \tfrac{1}{2} a^2 (\td\theta^2 +\sin^2\theta \td\phi^2)   + a^2 \sigma_3'^2  \; .
\ee
Furthermore,  we may compute the first order data.  In particular, we need $\gamma^{(1)}_{ab} = \partial_\lambda \gamma_{ab} |_{\lambda=0}$. We find
\be
\gamma^{(1)}_{ab}dx^a dx^b =\left(  \frac{dr}{d\lambda} \;   \partial_r\gamma_{ab} \right)_{r= \sqrt{2} a} \td x^a \td x^b =  a (\td\theta^2 +\sin^2 \theta \td\phi^2)  \; .
\ee
Note that the $\sigma_3'^2$ term is absent since $ \partial_r h^2 |_{r= \sqrt{2} a}=0$.

\subsection{Extreme KK black hole}

The extreme Kaluza-Klein black hole with zero angular momentum is the \textit{two}-parameter family of solutions~\cite{Larsen:1999pp},
\begin{equation}
\td s^2 = -\frac{r^2}{H_q} \; \td t ^2 + \frac{H_q}{H_p} \left( \td \psi - 2 P \cos \theta \; \td \phi -  \Omega  \td t \right)^2 + \frac{H_p}{r^2} \; \td r^2 + H_p ( \td \theta ^2 + \sin ^2 \theta \; \td \phi^2) 
\end{equation}
with 
\begin{eqnarray}
\nn & & H_p = r^2 + r p + \frac{p^2 q}{2 (p+q)}  \qquad  H_q = r^2 + r q + \frac{p q^2}{2 (p+q)} \qquad  \Omega = \frac{Q(2 r + p)}{H_q}  \\
& & P^2 = \frac{p^3}{4 (p+q)}  \qquad   Q^2 = \frac{q^3}{4 (p+q)} 
\end{eqnarray}
where the two parameters $p$ and $q$ are positive constants and they are related to the magnetic and electric charges. The surface $r=0$ is a smooth degenerate horizon.

Coordinate regular on the future horizon are given by
\be
\td v = \td t  + \frac{\sqrt{H_p H_q}}{r^2} \td r \qquad \td\psi' = \td\psi -  \Omega_H \td t - (\Omega-\Omega_H)\frac{\sqrt{H_p H_q}}{r^2}  \td r
\ee
in terms of which the metric is
\be
ds^2 = - \frac{r^2}{H_q} \td v^2 + 2 \sqrt{\frac{H_p}{H_q}} \td v \td r+ \frac{H_q}{H_p} \left( \td \psi' - 2 P \cos \theta \; \td \phi -  (\Omega-\Omega_H)  \td v \right)^2 +H_p ( \td \theta ^2 + \sin ^2 \theta \; \td \phi^2) \; .
\ee
To put this in Gaussian null coordinates we need to simply change radial variable to
\be
\lambda  = \int^r_0  \sqrt{\frac{H_p}{H_q}}  \td r   \; .
\ee
Near the horizon  $\lambda  = \sqrt{\frac{p}{q}} r + O(r^2)$. The coordinates $(v, \lambda, \theta, \phi, \psi')$ are Gaussian null coordinates and $x^a=(\theta, \phi, \psi')$ coordinates on a cross-section of the horizon. The horizon metric is easily read off,
\bea
\gamma_{ab}(x) \td x^a \td x^b &=& \frac{q}{p} (\td\psi'- 2P \cos \theta \td\phi)^2 + \frac{p^2 q}{2(p+q)} (\td\theta^2+\sin^2\theta \td\phi^2) \\
&=& \frac{p^2 q}{(p+q)} \left[ ( \td\psi''- \cos \theta \td\phi)^2 + \tfrac{1}{2}(\td\theta^2+\sin^2\theta \td\phi^2)  \right]
\eea
where in the second equality we have defined $\psi' = 2P \psi''$.

The first order data is now easily computed. We find
\be
\gamma^{(1)}_{ab} \td x^a \td x^b = \partial_\lambda \gamma_{ab}|_{\lambda=0} \td x^a \td x^b = \sqrt{pq} ( \td\theta^2 +\sin^2 \theta \td\phi^2) \; .
\ee
Note the $\psi'$ components vanish  due to $\partial_r (\frac{H_q}{H_p})|_{r=0}=0$.


\begin{thebibliography}{99}
{  \small

\bibitem{Emparan:2008eg} 
  R.~Emparan and H.~S.~Reall,
  Living Rev.\ Rel.\  {\bf 11} (2008) 6
  [arXiv:0801.3471 [hep-th]].
   
  \bibitem{Galloway:2005mf}
  G.~J.~Galloway and R.~Schoen,
  Commun.\ Math.\ Phys.\  {\bf 266} (2006) 571
  [gr-qc/0509107].

\bibitem{Hollands:2006rj}
  S.~Hollands, A.~Ishibashi and R.~M.~Wald,
  Commun.\ Math.\ Phys.\  {\bf 271} (2007) 699
  [gr-qc/0605106].
  
  \bibitem{Friedman:1993ty}
  J.~L.~Friedman, K.~Schleich and D.~M.~Witt,
  Phys.\ Rev.\ Lett.\  {\bf 71} (1993) 1486
   [Phys.\ Rev.\ Lett.\  {\bf 75} (1995) 1872]
  [gr-qc/9305017].
  
  \bibitem{Chrusciel:1994tr}
  P.~T.~Chrusciel and R.~M.~Wald,
  Class.\ Quant.\ Grav.\  {\bf 11} (1994) L147
  [gr-qc/9410004].
  
  \bibitem{Galloway}
  G.~J.~Galloway,
  Class. Quant. Grav. {\bf 12} (1995) L99-L101.
  
  \bibitem{Andersson:2015sfa}
  L.~Andersson, M.~Dahl, G.~J.~Galloway and D.~Pollack,
  arXiv:1508.01896 [gr-qc].
  
   \bibitem{Hollands:2012xy}
  S.~Hollands and A.~Ishibashi,
  Class.\ Quant.\ Grav.\  {\bf 29} (2012) 163001
  [arXiv:1206.1164 [gr-qc]].

  
  \bibitem{Hollands:2007aj}
  S.~Hollands and S.~Yazadjiev,
  Commun.\ Math.\ Phys.\  {\bf 283} (2008) 749
  [arXiv:0707.2775 [gr-qc]].
  
  \bibitem{Kunduri:2014iga}
  H.~K.~Kunduri and J.~Lucietti,
  JHEP {\bf 1410} (2014) 82
  [arXiv:1407.8002 [hep-th]].
  
  \bibitem{Kunduri:2014kja}
  H.~K.~Kunduri and J.~Lucietti,
  Phys.\ Rev.\ Lett.\  {\bf 113} (2014) 21,  211101
  [arXiv:1408.6083 [hep-th]].
  
\bibitem{Kunduri:2013gce}
  H.~K.~Kunduri and J.~Lucietti,
  Living Rev.\  Rel.\  {\bf 16} (2013) 8
  [arXiv:1306.2517 [hep-th]].
  
  \bibitem{Lewandowski:2002ua}
  J.~Lewandowski and T.~Pawlowski,
  Class.\ Quant.\ Grav.\  {\bf 20} (2003) 587
  doi:10.1088/0264-9381/20/4/303
  [gr-qc/0208032].
  
  \bibitem{Amsel:2009et}
  A.~J.~Amsel, G.~T.~Horowitz, D.~Marolf and M.~M.~Roberts,
  Phys.\ Rev.\ D {\bf 81} (2010) 024033
  [arXiv:0906.2367 [gr-qc]].
  
  \bibitem{Figueras:2009ci}
  P.~Figueras and J.~Lucietti,
  Class.\ Quant.\ Grav.\  {\bf 27} (2010) 095001
  [arXiv:0906.5565 [hep-th]].
  
  \bibitem{Chrusciel:2010gq}
  P.~T.~Chrusciel and L.~Nguyen,
  Annales Henri Poincare {\bf 11} (2010) 585
  [arXiv:1002.1737 [gr-qc]].
  
  \bibitem{Meinel:2011wu}
  R.~Meinel,
  Class.\ Quant.\ Grav.\  {\bf 29} (2012) 035004
  [arXiv:1108.4854 [gr-qc]].
  
  \bibitem{Kunduri:2007vf}
  H.~K.~Kunduri, J.~Lucietti and H.~S.~Reall,
  Class.\ Quant.\ Grav.\  {\bf 24} (2007) 4169
  [arXiv:0705.4214 [hep-th]].
  
  
   
   \bibitem{Reall:2002bh}
  H.~S.~Reall,
  Phys.\ Rev.\ D {\bf 68} (2003) 024024
   [Phys.\ Rev.\ D {\bf 70} (2004) 089902]
  [hep-th/0211290].
  
  \bibitem{Kunduri:2008rs}
  H.~K.~Kunduri and J.~Lucietti,
  J.\ Math.\ Phys.\  {\bf 50} (2009) 082502
  [arXiv:0806.2051 [hep-th]].


 \bibitem{Amsel:2009ev}
  A.~J.~Amsel, G.~T.~Horowitz, D.~Marolf and M.~M.~Roberts,
  JHEP {\bf 0909} (2009) 044
  [arXiv:0906.2376 [hep-th]].
  
  \bibitem{Dias:2009ex}
  O.~J.~C.~Dias, H.~S.~Reall and J.~E.~Santos,
  JHEP {\bf 0908} (2009) 101
  [arXiv:0906.2380 [hep-th]].
  
  
  
  \bibitem{Durkee:2010ea}
  M.~Durkee and H.~S.~Reall,
  Phys.\ Rev.\ D {\bf 83} (2011) 104044
  [arXiv:1012.4805 [hep-th]].
  
  \bibitem{Hollands:2014lra}
  S.~Hollands and A.~Ishibashi,
  Commun.\ Math.\ Phys.\  {\bf 339} (2015) 3,  949
  [arXiv:1408.0801 [hep-th]].

 \bibitem{Booth:2012xm}
  I.~Booth,
  Phys.\ Rev.\ D {\bf 87} (2013) 2,  024008
  [arXiv:1207.6955 [gr-qc]].

  
  \bibitem{Li:2013pra}
  C.~Li and J.~Lucietti,
  Phys.\ Lett.\ B {\bf 738} (2014) 48
  [arXiv:1312.2626 [hep-th]].
  
  \bibitem{Booth:2005qc}
  I.~Booth,
  Can.\ J.\ Phys.\  {\bf 83} (2005) 1073
  [gr-qc/0508107].
  
  \bibitem{Lucietti:2012sf}
  J.~Lucietti and H.~S.~Reall,
  Phys.\ Rev.\ D {\bf 86} (2012) 104030
  [arXiv:1208.1437 [gr-qc]].

\bibitem{Andersson:2007fh}
  L.~Andersson, M.~Mars and W.~Simon,
  Adv.\ Theor.\ Math.\ Phys.\  {\bf 12} (2008) 853
  [arXiv:0704.2889 [gr-qc]].

\bibitem{Booth:2007wu}
  I.~Booth and S.~Fairhurst,
  Phys.\ Rev.\ D {\bf 77} (2008) 084005
  [arXiv:0708.2209 [gr-qc]].
  
  \bibitem{Mars:2012sb}
  M.~Mars,
  Class.\ Quant.\ Grav.\  {\bf 29} (2012) 145019
  [arXiv:1205.1724 [gr-qc]].
  
  \bibitem{Hayward:1993wb}
  S.~A.~Hayward,
  Phys.\ Rev.\ D {\bf 49} (1994) 6467.
  
  \bibitem{Racz:2008tf}
  I.~Racz,
  Class.\ Quant.\ Grav.\  {\bf 25} (2008) 162001
  [arXiv:0806.4373 [gr-qc]].
  
   
  \bibitem{Figueras:2008qh}
  P.~Figueras, H.~K.~Kunduri, J.~Lucietti and M.~Rangamani,
  Phys.\ Rev.\ D {\bf 78} (2008) 044042
  [arXiv:0803.2998 [hep-th]].
  
  \bibitem{Lucietti:2012sa}
  J.~Lucietti,
  Class.\ Quant.\ Grav.\  {\bf 29} (2012) 235014
  [arXiv:1209.4042 [gr-qc]].
  
  \bibitem{Li:2013gca}
  C.~Li and J.~Lucietti,
  Class.\ Quant.\ Grav.\  {\bf 30} (2013) 095017
  [arXiv:1302.4616 [hep-th]].
  
  
   \bibitem{Moncrief:1983xua}
  V.~Moncrief and J.~Isenberg,
  Commun.\ Math.\ Phys.\  {\bf 89} (1983) 3,  387.

  
  \bibitem{Booth:2015nwa}
  I.~Booth, M.~Hunt, A.~Palomo-Lozano and H.~K.~Kunduri,
  arXiv:1502.07388 [gr-qc].
  
  \bibitem{Kunduri:2006qa}
  H.~K.~Kunduri, J.~Lucietti and H.~S.~Reall,
  Phys.\ Rev.\ D {\bf 74} (2006) 084021
  [hep-th/0606076].
  
  \bibitem{Dias:2015rxy}
  O.~J.~C.~Dias, J.~E.~Santos and B.~Way,
  arXiv:1505.04793 [hep-th].
  
  \bibitem{Larsen:1999pp}
  F.~Larsen,
  Nucl.\ Phys.\ B {\bf 575} (2000) 211
  [hep-th/9909102].
  
}
\end{thebibliography}
\end{document}